\documentclass[11pt]{article} 
\usepackage[utf8]{inputenc}
\usepackage{multirow}
\usepackage{amsfonts}
\usepackage{amsmath}
\usepackage{amssymb}
\usepackage{amsthm}
\usepackage{caption}
\usepackage[dvipsnames]{xcolor}
    \definecolor{darkgreen}{rgb}{0,0.5,0}
    \definecolor{darkblue}{rgb}{0,0,0.6}
    \definecolor{purple}{rgb}{0.4,.2,0.7}
\usepackage[margin = 2.5cm]{geometry}
\usepackage{graphicx}
\usepackage[hyperfootnotes = false, colorlinks = true, linkcolor = darkblue, citecolor = purple]{hyperref}
\usepackage{subcaption}
\usepackage{ytableau}

\usepackage{comment}

\newcommand{\be}{\begin{equation}}
\newcommand{\ee}{\end{equation}}
\newcommand{\bea}{\begin{eqnarray}}
\newcommand{\eea}{\end{eqnarray}}
\def\la{\label}
\def\nref#1{(\ref{#1})}
\def\half{{1 \over 2 }}

\newcommand{\JM}[1]{{ \color{blue} #1 - JM }}
\newcommand{\YC}[1]{{ \color{darkgreen} #1 - YC }}
\newcommand{\VI}[1]{{ \color{red} #1 - VI }}

	\newcommand{\bes}{\begin{equation} \begin{split} }	
	\newcommand{\ees}{\end{split} \end{equation} }

	\usepackage{physics}

\begin{document}

\thispagestyle{empty}
\begin{center}
    ~\vspace{5mm}

  \vskip 2cm 
  
   {\LARGE \bf 
       Comments on the double cone wormhole 
   }

   \vspace{0.5in}
     
   {\bf   Yiming Chen$^{\bowtie,\rhd}$, Victor Ivo$^\rhd$ and  Juan Maldacena$^\lhd$ 
   }

    \vspace{0.5in}
 
  $^{\bowtie} $
  Stanford Institute for Theoretical Physics, Stanford, CA 94305, USA 
   \\
   ~
   \\
  $^\rhd$
  Jadwin Hall, Princeton University,  Princeton, NJ 08540, USA 
   \\
   ~
   \\
  $^\lhd$
  Institute for Advanced Study,  Princeton, NJ 08540, USA

    \vspace{0.5in}

    \vspace{0.5in}
    

\end{center}

\vspace{0.5in}

\begin{abstract}
 
 In this paper we revisit the double cone wormhole introduced by Saad, Shenker and Stanford (SSS), which was shown to reproduce the ramp in the spectral form factor. As a first approximation we can say that this solution computes $\textrm{Tr}[ e^{ -i K T} ]$, a trace of the ``evolution'' operator that generates Schwarzschild time translations on the two sided wormhole geometry. This point of view leads to a simple way to compute the normalization factor of the wormhole. When we have bulk matter fields,  SSS suggested using a modified evolution $\tilde K$ which involves a slightly complex geometry, so that we are really computing $\textrm{Tr}[ e^{ - i \tilde K T} ]$. We argue that, for general black holes, the spectrum of $\tilde K$ is given by quasinormal mode frequencies. We explain that this reproduces various features that were previously predicted from the spectral form factor on hydrodynamics grounds. We also give a general algebraic construction of the modified boost in terms of operators constructed from half sided modular inclusions. For the special case of JT gravity, we work out the backreaction of matter on the geometry of the double cone and find that it deforms the geometry in an undesirable direction. We finally give some comments on the possible physical interpretation of $\tilde K$. 
   
 \end{abstract}
 
\vspace{1in}

\pagebreak

\setcounter{tocdepth}{3}
{\hypersetup{linkcolor=black}\tableofcontents}

\section{Introduction and motivation }

\begin{figure}[h]
    \begin{center}
   \includegraphics[scale=.35]{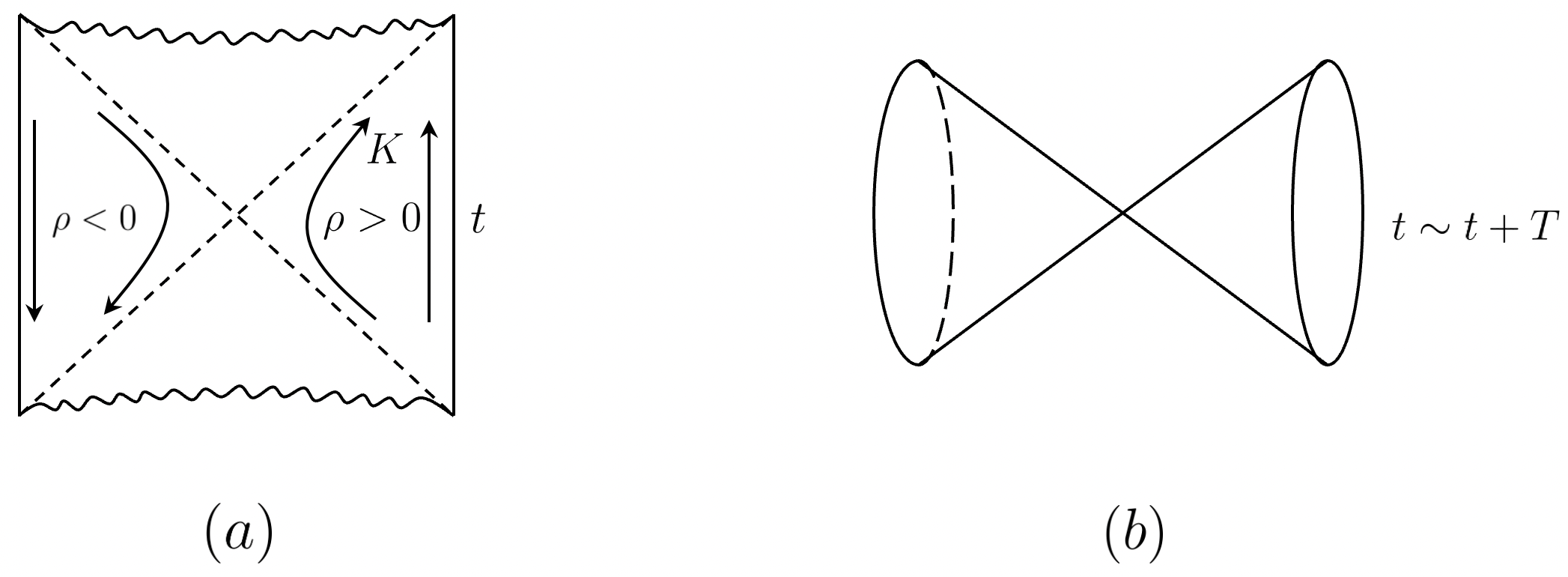}
    \end{center}
    \caption{(a) The geometry of a two-sided black hole. We have an symmetry $K=i\partial_t$ that acts like a boost near the horizon and like a time translation far away. (b) The double cone geometry involves quotienting the two-sided black hole such that $t$ has period $T$.   }
    \label{fig:penrose}
\end{figure}

Saad, Shenker and Stanford have introduced a very interesting and general spacetime wormhole solution \cite{Saad:2018bqo}. It can be constructed whenever we have a finite temperature black hole solution. The full extended black hole spacetime has two exterior regions and the geometry has a symmetry $K= i \partial_t$ that looks like a boost in the near horizon region and like a time translation symmetry far away from the black hole horizons, see fig. \ref{fig:penrose} (a). 
The construction in  \cite{Saad:2018bqo} involves a compactification of the time coordinate associated to $K$, $t \sim t + T$. This identification leads to a double cone wormhole, see fig. \ref{fig:penrose} (b).
 
From the bulk point of view, the double cone wormhole is essentially computing the trace of the operator $e^{ - i K T }$ in the  bulk Hilbert space  defined on a constant time slice. The trace sums over different Einstein-Rosen bridge configurations as well as the states of the quantum fields propagating on them.\footnote{Notice that this trace is over the type I algebra of the wormhole which includes the two sides. This should not be confused with traces over the type II algebras discussed in \cite{Witten:2021unn,Chandrasekaran:2022eqq}.} However, the trace is a bit singular for the quantum fields due to the degenerating circle at the center of the wormhole. \cite{Saad:2018bqo} introduced a certain $i \epsilon$ prescription that defines a complex geometry which leads to a new operator $\tilde K$, for which the  trace $\textrm{Tr}[e^{-i\tilde{K}T}]$ is well-defined and can be computed. We will call this new operator $\tilde{K}$ the ``modified boost operator".   

The goal of this paper will be to elaborate on the interpretation of the double cone wormhole as computing a trace, as well as understanding the properties of the modified boost operator $\tilde{K}$. At a basic level, this is useful in understanding the perturbative quantum fluctuations around the wormhole in \cite{Saad:2018bqo}, which is essential in understanding the role of such wormholes in gravity theories in higher dimensions.\footnote{See also \cite{Cotler:2021cqa,Mahajan:2021maz,Cotler:2022rud} for work on understanding different aspects of the double cone wormhole. 
}  Another reason to study this modified boost operator further is that it appears in various other computations 
\cite{Stanford:2020wkf,Bah:2022uyz}. The long time exponential of this modified boost operator, $ e^{ - i  \tilde{K} t}$, is  a non-unitary operator that eventually projects us onto the thermofield double state, or, more precisely, onto the thermofield double subspace which is classically described by the ADM mass and the relative time shift coordinate. 
Part of our motivation was also that a better bulk understanding might help us elucidate the boundary theory interpretation of this solution, in particular the reason that it is not factorizing into a product of two partition functions  
 \cite{Saad:2018bqo}.
 



The structure is the following.

In Section \ref{sec:complex} we review the geometry of the wormhole and the $i\epsilon$ prescription by Saad-Shenker-Stanford.

In Section~\ref{sec:trace}, we point out that viewing the construction as a trace helps us determine the overall normalization coefficient of the wormhole, which is important for the dual quantum mechanical interpretation of this configuration in terms of the spectral form factor  \cite{Cotler:2016fpe,Saad:2018bqo}. 

In Section~\ref{sec:mod}, we also explore some aspects of the modified boost evolution $\tilde K$ defined in \cite{Saad:2018bqo} on general black hole backgrounds. We review the result in \cite{Bah:2022uyz} that the eigenvalues of $\tilde K$ correspond to the quasinormal mode frequencies $\omega_{\textrm{QNM}}$ of the black hole. This is interesting because it gives the quasinormal modes as the eigenvalues of a (non-Hermitian) operator.  A consequence of this is that the one-loop determinant of (bosonic) quantum fields on the double cone takes the following universal form\footnote{If we have fermionic fields, then there are similar factors in the numerator from the fermions, see (\ref{Zoneloop}). }
\begin{equation} \la{DetQNM}
	Z_{\textrm{one-loop}} (T)  = \prod_{\omega_{\textrm{QNM}}} \frac{1}{1-e^{-i\omega_{\textrm{QNM}} T}}.
\end{equation}
We provide some simple examples of this general expression.

In Section~\ref{sec:algebra}, we also present a slightly more abstract way of thinking about the definition of the modified boost operator $\tilde K$ in terms of the symmetries of the near horizon geometry, by drawing a connection to the large $N$ von Neumann algebra as discussed in \cite{Leutheusser:2021frk}. This can be used to justify some of the properties of the modified boost in general situations. 

In Section \ref{sec:backreaction}, we study the effect of backreaction from the matter on the double cone wormhole. We find that this produces a complex deformation with the ``wrong sign''. We also discuss the case when we deform the problem to complex couplings which gives a deformation of the geometry with the right sign. We discuss some interpretations of these results. 

In Section~\ref{sec:toy}, in order to help understand the difference between the ordinary boost operator $K$ and the modified boost operator $\tilde{K}$ better, we also give a simple toy model in a harmonic oscillator Hilbert space that contains  operators similar to  $K$ and $\tilde K$. The toy model is helpful for understanding the relation between the two operators and the properties of their eigenvalues and eigenfunctions.

\section{The two-sided black hole geometry and its complexification } \label{sec:complex}


In this paper, we consider a general finite temperature black hole with the metric 
  \be \la{BHme}
  ds^2 =  - f(\rho) dt^2  + d\rho^2 + ds^2_\Sigma ~,
  \ee 
  where $\Sigma$ can be any space. Near the horizon we assume that $f(\rho) \propto \rho^2 + \cdots $, where the dots are higher order terms in the small $\rho$ expansion. 
  The two sides of the black hole have $\rho>0$ (right side) and $\rho<0$ (left side), see the Penrose diagram in figure \ref{fig:penrose} (a). This can describe general finite temperature black holes such as Schwarzschild $AdS_D$ black holes in any number of dimensions. With some small modifications we expect to be able to describe also rotating black holes, but we will not discuss that here. We will further assume that the Hartle Hawking state exists. 
  
  Shifts in the coordinate $t$ define a symmetry of this metric denoted by $K$. We see that $K$ acts like a boost symmetry near $\rho \sim 0$. 
  
  It is possible to define a Hilbert space around this solution. A first step is to notice that there is a family of solutions characterized by the mass $M$ and the relative time shift $T_{\textrm{rel}}$ between the two sides. In other words, the physical time $t$ on the two sides is given by 
  \be 
 t_{\rm phys}^L =   t ~,~~~~~~~~~~t_{\rm phys}^R =  t  + T_{\textrm{rel}} \la{TimeSh}
   \ee 
   We can think of these physical times as the boundary times on the two sides. 
   These two classical variables obey the Poisson bracket \cite{Kuchar:1994zk} (see \cite{Harlow:2018tqv} for a discussion in JT gravity)
   \be 
   \label{canonconj}
   \{  T_{\textrm{rel}}, M \} = 1 
   \ee 
   so that a simple semiclassical quantization would make them canonically conjugate operators.

 If there are gauge symmetries (discrete or continuous) there will be discrete gauge field holonomies or Wilson lines, extending from the left to the right boundary, that also characterize the configuration. This means we will need to sum or integrate over these.   In particular, these are present if $\Sigma$ has isometries, which can be viewed as giving rise to Kaluza-Klein gauge fields in the two dimensional geometry spanned by $t,\rho$. 
 
 As is completely standard, on this two sided geometry we can also define a Hilbert space for quantum fields.   We can choose the Hartle-Hawking vacuum state which is invariant under $K$, and we can add perturbations by acting with quantum fields on this state. 
 
\begin{figure}[h]
    \begin{center}
   \includegraphics[scale=.35]{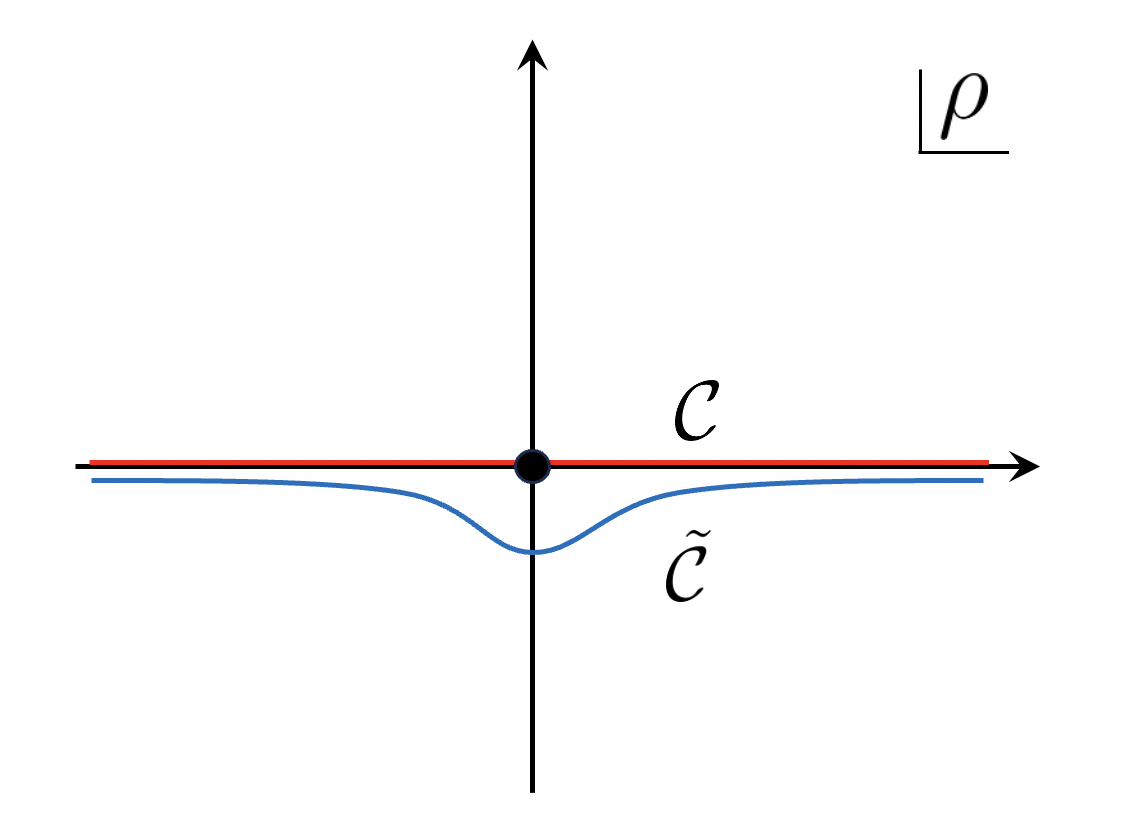}
    \end{center}
    \caption{SSS \cite{Saad:2018bqo} proposed a prescription to regulate the double cone geometry, by deforming the slice of $\rho$ on which the path integral for quantum fields is defined from the real axis $\mathcal{C}$ to $\tilde{\mathcal{C}}$ that has a tiny excursion into the lower half plane.}
    \label{fig:contour}
    \end{figure}

 
 As instructed by \cite{Saad:2018bqo}, we are also interested in a certain complex deformation of the geometry that is defined as follows. In the original spacetime (\ref{BHme}), $\rho$ takes real value from $-\infty$ to $\infty$. However, we could in general consider complexifying (\ref{BHme}), in particular, allow $\rho$ to be complex. In the complex plane of $\rho$, the real axis is a specific contour from $-\infty$ to $\infty$, which we denote by $\mathcal{C}$. 
 Now, for arbitrary time $t$, we   deform the contour   
 for $\rho$ as
    \be \la{AnalC} 
\rho =  \tilde \rho - i \epsilon , \quad \epsilon >0
\ee 
near $\rho \sim 0$ with real $\tilde \rho$.\footnote{In fact, \cite{Saad:2018bqo}   defined the deformation with an opposite sign, namely $\epsilon<0$. This is due to an opposite sign in their convention compared to us - in their convention, the double cone computes $\textrm{Tr}[e^{iKT}]$. The importance of the sign of $\epsilon$ will be clear from the general discussion in section \ref{sec:algebra}.       }    We denote this new contour as $\tilde{\mathcal{C}}$, see fig. \ref{fig:contour}. As we will see in section \ref{sec:mod}, the precise form of this deformation away from the horizon does not play an  important role. 
Notice also that since the spatial manifold $\Sigma$ remains finite size everywhere in the geometry, we do not need to consider the effect of this infinitesimal deformation on it. 
Since the new contour represents just a (complex) coordinate transformation of the solution, it it still a solution of the original gravity equations and its gravitational action is unchanged. 
   
 A different (but equivalent) perspective, which we will often adopt, is that we are fixing $\tilde \rho$ to be real, but rather changing the real metric (\ref{BHme}) (with $\rho$ there substituted by $\tilde{\rho}$) to be slightly complex 
   \be \la{NHmec}
  ds^2 \sim -  (\tilde \rho - i \epsilon)^2 dt^2 + d\tilde \rho^2 
  \ee 
  for real $\tilde \rho$ in the $\tilde \rho \sim 0$ region. 
  By definition, the operator $\tilde K$ is the operator that generates time translation in the complex metric (\ref{NHmec}). This  operator is non-Hermitian, a fact that will become clearer later. 
  This perspective makes it clear at what stage the difference between $K$ and $\tilde{K}$ comes in. We can first consider a constant time slice, described by $d\tilde \rho^2 + ds_{\Sigma}^2$, and construct the Hilbert space of quantum fields living on it. The $i\epsilon$ deformation doesn't show up in this step, which makes it clear that we have a \emph{single} Hilbert space. The $i\epsilon$ deformation only comes in when we discuss the time evolution or the trace, where we need to decide whether we are evolving between different slices with $K$ or $\tilde{K}$. Notice that this evolution is a boost-like evolution.   


\section{The wormhole as a trace and the normalization factor }
\label{sec:trace}
  
   The authors of \cite{Saad:2018bqo} constructed a wormhole that involved a periodic identification of the time $t$. 
 They argued that this wormhole is an approximation to the spectral form factor $ \overline{|Z(\beta + i T) |^2}$ of the quantum mechanical system that describes the black hole, suitably averaged over couplings and for  a suitable range of parameters where it displays a ``ramp''. 
More precisely, it was found that it is a bit better to think in terms of a modified spectral form factor where the sum over states is weighted by a ``window function'' $f(E)$ focusing on some range of energies. For example, this could be a simple Gaussian function, which was the case considered in \cite{Saad:2018bqo}. Then we compute 
 \be \label{Zf}
  \overline{ |Z_f (T)|^2 } ~,~~~~~~~{\rm with} ~~~~~~~ Z_f (T)\equiv \textrm{Tr}_{\textrm{bdy}}[ f(H_b) e^{ - i H_bT } ]~,
\ee 
where $H_b$ is the Hamiltonian of the quantum mechanical system describing the black hole.\footnote{Not to be confused with the operator $H$ appearing in the bulk discussion in the latter sections.} We added a subscript ``bdy" to the trace to highlight it is a trace taken in the boundary exact description. All the traces without the explicit subscript are taken in the bulk effective theory.

This  interpretation implies that the double cone wormhole is supposed to compute 
 \be \la{WHEx}
Z_{\textrm{dc}}  \sim  \int d E d E' \rho_c(E,E') e^{ - i (E-E') T} f(E) f(E')      
  \ee 
 where the subscript ``dc" stands for double cone.
The function   $f(E)$  should have a width much larger than $1/T$. We are also in the regime where   $1/T$ is larger than the energy spacing between consecutive eigenvalues, meaning that $T$ is not too large.  In this regime, the connected pair correlation function of eigenvalues, which arises due to the average, is given by \cite{Mehta} 
  \be \la{PairDis}
   \rho_c(E,E')= - { 1 \over 2 \pi^2 (E-E')^2 }  
  \ee 
  where we have assumed that the system has no symmetries and belongs to the gaussian unitary ensemble.\footnote{For relativistic theories that have time reversal and spatial reflection symmetries, we should have an extra factor of two in (\ref{PairDis}) and some of the following. See section 4 of \cite{Yan:2023rjh}.  } This formula is valid when $|E-E'|$ is significantly larger that the typical spacing between eigenvalues.   An important point about \nref{PairDis} is that the normalization is independent of the details of the system.  Therefore we expect that this normalization should be reproduced by the gravity computation. This was indeed checked in \cite{Saad:2018bqo} for some cases. 
  We would like to give a general argument that is valid for any black hole. 
  
  First we note that when we insert \nref{PairDis} into \nref{WHEx} and integrate over $E-E'$ we get 
  \be \la{CanWh}
   \int         d E  \, \,{ |T | \over 2 \pi } \times f(E)^2= 
   \int dE \int_0^{|T|} dT' \, {  1 \over 2 \pi }  \times f(E)^2 .
  \ee 
  The goal is to explain this from gravity, including the prefactor.  
  In deriving this formula we have assumed that the function $f(E)$ is close to constant over a scale of energies up to $1/T$. In other words, if we pick a gaussian 
  \be \la{ExpfDe}
  f(E) \propto \exp\left[ - \half { (E-E_0)^2 \over \Delta_E^2 } \right],
   \ee 
   then we require 
   \be \la{DeltaAp}
   \Delta_E T \gg 1.
   \ee 
   
 \subsection{General gravity computation } \label{sec:genegrav}
 
  We can understand the specific normalization in (\ref{CanWh}) from the gravity point of view as follows. We interpret the gravity solution as a trace over two sided configurations  
    \be \la{Zgr}
 Z_{\textrm{dc}} \sim   \textrm{Tr}[  f(E)^2 e^{ - i   \tilde K   T } ] ~, 
   \ee 
   Here $  \tilde K$ is the generator of the boost-like isometries for the geometry deformed as in \nref{AnalC}. 
   For what we will discuss now, we can replace $\tilde K$ by $K$ in this formula. The difference between them will be discussed in sec. \ref{sec:mod}.
   
   More precisely, the double cone wormhole involves a compactification of both the left and right times 
   $t_L \sim t_L + T$ and $t_R \sim t_R+T$. This implies that we have a compactification of the boost time shift variable discussed in \nref{TimeSh}, 
   \be 
   \la{Tshift} 
   T_{\textrm{rel}} \sim T_{\textrm{rel}} + T.
   \ee    
When we take the trace, we will have to sum over the family of classical wormhole configurations parameterized by $E$ and $T_{\textrm{rel}}$. Since they are cannonically conjugate variables, this means that the trace can be approximated by an integral over these variables with the standard measure factor 
     \be 
    { d E d T_{\textrm{rel}} \over 2 \pi } 
    \ee 
 We now see that this reproduces the expectation from the spectral form factor \nref{CanWh}. 
 In other words, we have 
 \be \la{WhTr} 
\textrm{Tr}[ f(E_L) f(E_R) e^{ - i K T} ]_T = \textrm{Tr}[ f(E)^2 ]_T =    \int dE \int_0^{|T|} dT_{\textrm{rel}} \, {  1 \over 2 \pi }  \times f(E)^2
\ee 
where the left and right energies are the same and we used that $K$ annihilates the empty wormholes. In addition, the subscript $ T$ emphasizes that  that the relative time shift coordinate $T_{\textrm{rel}}$ has also been identified as in \nref{Tshift}. So, $T$ appears in the definition of the phase space of wormhole configurations. 
In deriving \nref{WhTr} we ignored the quantization of $E$ that would be implied by making $T_{\textrm{rel}}$ periodic which is reasonable in the approximation \nref{DeltaAp}. We discuss the issue of compactifying $T_{\rm rel}$ further in sec. \ref{sec:algebratrace}.  
 
 The formula \nref{WhTr} was independently derived by \cite{DouglasPrivate}. 
  
  In addition, we need to take the trace over the quantum fields. This trace over the quantum fields was computed in \cite{Saad:2018bqo} for the special case of a near extremal black hole, where the geometry can be approximated by $AdS_2$. In the next section, we will discuss this trace in general and argue that for large values of $T$, it becomes one, thus not modifying the linear ramp at late time.  
  
  More precisely, we are saying that the full formula for the double cone wormhole partition function is 
  \be 
  Z_{\rm dc} \sim   \left( { |T| \over 2 \pi} \int dE   f(E)^2 \right) 
\times {\rm Tr}[ e^{ - i  \tilde K  T} ] 
\ee 
where the last trace is over the bulk quantum field theory Hilbert space, ignoring the mass and relative timeshift variables. We will now describe the computation of the last factor.

 \section{The modified boost operator $\tilde{K}$}\label{sec:mod}

  In this section, we consider a general finite temperature black hole with the metric 
 \nref{BHme},  but we now set the temperature to $\beta = 2\pi $. It is easy to restore factors of the temperature later. We consider the complex metric \nref{NHmec}, which results from setting $\rho = \tilde \rho - i \epsilon$ in \nref{BHme}. So, when we solve the wave equation we can use the variable $\rho$, but when we join the solutions from the left to the right we need to remember they are connected through the variable $\tilde \rho$. $\tilde K$ is the operator evolving the fields with this complex metric.

  Using this definition of $\tilde K$ we can proceed to find its eigenvectors and eigenvalues.\footnote{This argument was originally given in an appendix in \cite{Bah:2022uyz}. We reproduce it here for convenience.} 
  We consider fields propagating on the geometry \nref{BHme}. After decomposing them into eigenfunctions of the space $\Sigma$ we have a two dimensional problem. We can say that the field has a definite eigenfrequency   
$\phi(t,\rho) = e^{-i \omega t} \phi_\omega(\rho)$, where $\phi_\omega(\rho)$ obeys the wave equation 
\be \la{Weq}
  { 1 \over \sqrt{f(\rho)  g_{_\Sigma  }}} \partial_\rho \left[ \sqrt{f(\rho)  g_{_\Sigma}} \partial_\rho \phi_\omega \right] + { \omega^2 \over f } \phi_\omega - m_{\textrm{eff}}^2 \phi_\omega = 0 
\ee 
where $m_{\textrm{eff}}^2 $ is the effective two dimensional mass after taking into account the higher dimensional mass as well as the eigenvalue of the wave equation in the $\Sigma $ space.\footnote{We defined $ m_{\textrm{eff}}^2 = m^2  - { 1 \over \sqrt{ g_{_\Sigma} } } \partial_{\alpha } [ g^{\alpha \beta } _{_\Sigma}\sqrt{g_{_\Sigma }} \partial_{\beta } ] $ and we are imagining that we diagonalize the last operator. It could generally be a function of $\rho$.} We imagine that we have an AdS boundary condition at large $\rho$ which selects one of the two independent solutions of \nref{Weq}. Let us denote that solution as $D_\omega(\rho)$. It is defined up to an overall normalization which we will not fix. It is the solution that obeys the proper boundary condition on the right side, $\rho>0$. There is a similar solution that obeys the proper boundary condition on the left side. It is given by $D_\omega( -\rho)$ for $\rho<0$.  

The two independent solutions near $\rho\sim 0$ behave as $\rho^{ \pm i \omega }$. This means that the solution we constructed behaves as 
 \begin{equation} \la{Rsid}
D_{\omega}(\rho)  \sim a({\omega})\rho^{i \omega}+b(\omega) \rho^{-i \omega} ~,~~~~\rho>0 .
\end{equation}
Note that the boundary conditions set the ratio of $a(\omega)/b(\omega)$. 
We have a similar expansion for the left side solution 
\begin{equation} \la{Lsid}
D_{\omega}(-\rho)  \sim a({\omega})( -\rho)^{i \omega}+b(\omega) (-\rho)^{-i \omega}~,~~~~\rho<0. 
\end{equation}
Now, we have the requirement that the solution is analytic when we continue it through the lower half plane, $\rho \to e^{ - i \pi } \rho$, which is the same as going from positive to negative $\tilde \rho$ along the real $\tilde \rho $ line, see fig. \ref{fig:contour}.  This means that the analytic continuation of \nref{Rsid} is 
 \begin{equation} \la{Rcont}
D_{\omega}(e^{ - i \pi } \rho)  \sim a({\omega}) e^{  \pi \omega} \rho^{i \omega}+b(\omega) e^{ -\pi \omega } \rho^{-i \omega} = \Lambda(\omega) D_\omega(\rho)  
\end{equation}
where we equated this continuation with the left hand side  \nref{Lsid} up to an overall constant $\Lambda(\omega)$, since the solutions on each side are determined up to an overall constant. In particular, this implies that the ratio of the two coefficients of $\rho^{\pm i \omega}$ should remain the same. There are three ways this can be true:    $a({\omega})=0$,   $b({\omega})=0$ or $\omega=  i n$, where $n$ an integer.  

The condition  
\be \la{Azer}
 a(\omega)=0  
 \ee 
gives a quasinormal mode. These are modes that are non-singular in the future horizon. 
 To see this, it is useful to go to flat space coordinates near the horizon $X^\pm = \pm \rho e^{ \pm t }$.   The full  wavefunction \nref{Rsid} then behaves as 
 \be \la{SolHor}
 D_\omega (\rho)e^{ - i \omega t } \sim  a(\omega) (-X^-)^{i \omega } + b(\omega) (X^+)^{ - i \omega } .
 \ee 
  The future horizon is at $X^-=0$. The term involving $a(\omega)$ is rapidly oscillating and non-analytic  there. The solution becomes smooth at the future horizon if we set the condition \nref{Azer}, which indeed defines the quasinormal modes. 
  
  The condition $b(\omega)=0$ defines anti-quasinormal modes which are regular in the past horizon, $X^+=0$. In a stable background, all solutions to \nref{Azer}   have a negative imaginary part. 
The solutions of $b(\omega)=0$ are opposite in sign and have a positive imaginary part.  We will later argue in section \ref{sec:algebra} that as a more abstract definition of the quantum operator $\tilde K$ makes it clear that it has non-positive imaginary part, so that only quasinormal modes can be proper eigenvalues of  $\tilde K$. We also present in appendix \ref{subsec:kscrit} a different point of view on the imaginary part of $\tilde K$. Of course, by just solving the wave equation we get solutions with both signs.
  
  Finally, if we instead assumed that $a\not =0$, $b\not =0$ and   $\omega=  i n$, then  
 the two indices of the radial equation would differ by an integer. Generically this means that  the   most divergent solution near $\rho \sim 0$   contains a logarithmic term, that is:
\begin{equation}
   \phi_{1} \sim \rho^{-|n|}+....+ c    \rho^{|n|}  \log \rho  + \cdots     ~,~~~~~~~~~\phi_2 \sim \rho^{|n|}  + \cdots  
\end{equation} 
This logarithm   gives an extra term when we go from the right to the left and will not   match the solution with the right boundary conditions.  Generically we expect that   the coeffcient of the log term  does not vanish,  so that the mode will not obey the SSS boundary conditions.  
In the very special case when the coefficient of the logarithm vanishes, then we see from \nref{SolHor} that when $n< 0$ the mode is   smooth at the future horizon, so it is also a quasinormal mode.\footnote{This special case arises for a massless field in $AdS_2$ as discussed in \cite{Saad:2018bqo}, as we   review in section \ref{QNMExpl}.}  Conversely, if the coefficient of the logarithm is non-zero, then the solution will not be smooth at the future horizon and will not be a quasinormal mode.

  So, we conclude that the quasinormal modes are right eigenvectors of the operator $\tilde K$ defined by doing evolution in the complex geometry defined by \nref{BHme} and \nref{NHmec}. We can summarize this result through the simple eigenvalue equation for $\tilde{K}$
\begin{equation}
	\tilde{K}\,\phi_{\textrm{QNM},n}=\omega_{\textrm{QNM},n}\,\phi_{\textrm{QNM},n}
\end{equation}
with $\omega_{\textrm{QNM},n}$ the $n$-th quasinormal mode of the system, and $\phi_{\textrm{QNM},n}$ its wavefunction.  Note that here we used ``right eigenvector" since $\tilde{K}$ is non-Hermitian, the right and left eigenvectors are typically different.  
Another remark is that the modified boost evolution operator $\tilde K$, defined through the complex metric, allows us to in principle define quasinormal modes beyond the free field approximation. $\tilde K$ is an operator defined in the second quantized theory and one could in principle include interactions. We can view the fields defined on this geometry as the ordinary Hilbert space of the two sided black hole, but undergoing evolution with a modified operator denoted by $\tilde K$. 

This means that the contribution from the fields propagating on the background takes the form 
\be \label{Zoneloop}
Z_{\rm one-loop}(T) = \textrm{Tr}[ e^{ - i  \tilde KT } ] = \prod_{\omega_{\rm QNM} } { 1 + e^{ - i \omega_f T } \over 1 - e^{ - i \omega_b T } } 
\ee 
where $\omega_f$ and $\omega_b$ are the quasinormal mode frequencies of the fermions and the bosons. Here the trace is just over the matter fields, the contribution from the unexcited wormhole states was taken into account in \nref{WhTr}.

If all quasinormal modes have finite negative imaginary part, then \nref{Zoneloop} goes to one as $T\to \infty$. However, some quasinormal modes might be fairly long lived, such as high angular momentum particles  for black holes in $AdS_d$, for $d\geq 4$, \cite{Dodelson:2022eiz} or hydrodynamic modes, see   section \ref{sec:hydro}, so we need to wait until a later time for them to be suppressed.  
One can also wonder how the backreaction from the one-loop determinant (\ref{Zoneloop}) modifies the double cone geometry. In section \ref{sec:backreaction}, we provide a study of the backreacted geometry in AdS$_2$.

In particular, note that we can define the Thouless time as the time where the random matrix behavior takes over, see \cite{Nosaka:2018iat,Winer:2020gdp}. This happens at a time $T $, defined to be the Thouless time,  when 
\nref{Zoneloop} becomes of order 1. In addition, we need that the disk contribution is subleading, for the exponential window function \nref{ExpfDe} this requires $ \Delta_E T \gtrsim \sqrt{S(\bar E) } $.

When we think about quasinormal modes for black holes in AdS, it is useful to think about the large mass limit \cite{Fidkowski:2003nf,Festuccia:2008zx}. In that case, one can use a semiclassical geodesic approximation to describe the quasinormal mode. The lowest quasinormal mode corresponds to a particle wordline extended along the $t$ direction and sitting a value of $\rho =\rho_*$ that extremizes its action 
\be 
 i I =-i m\int dt \sqrt{f(\rho_*)   } ~,~~~~~~~~~~ f'(\rho_*) =0 
 \ee 
 In general the value of $\rho_*$ is complex and we should choose it appropriately as the one that leads to the lowest   imaginary part of the right sign in $\sqrt{f(\rho_*)}$ \cite{Festuccia:2008zx}. 
This shows that the lowest quasinormal mode frequency is $\omega = m \sqrt{f(\rho_*)}$ and that we could think of   
\be 
 \ell =   \sqrt{f(\rho_*)}T ~,
 \ee 
 or its imaginary part, as the effective size of the throat of the wormhole. 

 This discussion presents the quasinormal modes as solutions of an eigenvalue problem for the non-hermitian operator $\tilde K$, the modified boost. In principle, this is an operator defined in the second quantized theory and we can also consider interactions.

 As a side comment, it was shown in \cite{Denef:2009kn} that the one-loop determinant, for a Euclidean black hole (instead of the double cone) also admits a representation 
in terms of the quasinormal modes.  In \cite{Law:2022zdq}  a trace   interpretation for this formula was proposed. It would be interesting to understand its relation to our discussion.

\subsection{Contributions from diffusive and hydrodynamic modes to the one loop determinant}

\la{sec:hydro}


In this section,  we discuss the contribution from the low frequency quasinormal modes associated to hydrodynamics, 
%
see 
\cite{Berti:2009kk} for a review. 
The contribution from each mode can be very easily read off from \nref{DetQNM}, and one can compute the  resulting hydrodynamic corrections to the ramp. 

As a simple example, consider a black brane and focus on the contribution from a diffusive mode with dispersion relation  
\be 
\omega = - i D q^2 
\ee
where $q$ (or really $\vec q$) is its spatial momentum. This can arise from a gauge field in the bulk and it represents the diffusion of charge in the boundary \cite{Policastro:2002se}. It can also arise from the shear mode fluctuations of the metric \cite{Kovtun:2005ev}, which are in the vector representation of the rotation group that preserves $\vec{q}$.  
When we take the trace, each diffusive mode gives  a contribution of the form
\begin{equation}
\label{TrDq}	
\text{Tr}\,[ e^{-i \tilde{K}T}]=
\prod_{\vec q} \frac{1}{1-e^{-D q^{2}T}}
\end{equation}
where each term in the product comes from the multiparticle contribution from each mode.\footnote{Here we focus on the contribution from non-zero modes. This corresponds to the answer where the two sides are in the same representation $R$ and have the same eigenvalues of the Cartan generators. The zero mode needs to be treated separately following the disccusion below  (\ref{canonconj}), which leads to an extra $\textrm{dim}(R)^2$ prefactor \cite{Kapec:2019ecr}. We thank Douglas Stanford for asking this.} 
 The   one loop contribution is then
\begin{equation}
\log \text{Tr}\, [e^{-i\tilde{K}T}]=
- V\int { d^{d}q \over (2\pi)^{d}} \log(1-e^{-D q^{2}T}) =\frac{V}{(4 \pi D T)^{\frac{d}{2}}}\zeta\bigg(1+\frac{d}{2}\bigg) 
\end{equation}
where $d$ is the number of spatial dimensions of the boundary theory. 
This  matches precisely the effective hydrodynamic description of the spectral form factor discussed in 
\cite{Winer:2020gdp}. As explained in \cite{Winer:2020gdp}, this factor can be viewed as the number of modes that has not decayed up to time $T$, so these are conserved quantities.  Here we see how this expected contribution arises very simply from the bulk theory. 
 Of course, this is not surprising once we know the general connection between gravity and hydrodynamics \cite{Berti:2009kk}, and the connection between hydrodynamics and the spectral form factor  \cite{Winer:2020gdp}.  

With this method we can also naturally include contributions from other hydrodynamic modes. In fact, the most relevant one at low energies is generally the sound mode, with dispersion relation
\begin{equation}
\label{soundpole}
\omega=\pm c_{s} |q|-i D q^{2}	
\end{equation}
with $c_{s}$ the speed of sound and $D$ the mode decay rate. The contribution is different depending on whether or not the number of dimensions $d$ is even or odd. For $d$ even we get 
\begin{equation}
\log \text{Tr}\, [e^{-i\tilde{K}T}]=V \int \frac{d^{d}q}{(2\pi)^{d}}\sum_{\eta=\pm}\sum_{n=1}^{\infty}\frac{e^{-i \eta n c_{s}|q|T-n D q^{2}T}}{n}=V\frac{4(-1)^{\frac{d}{2}}\Gamma(d)}{(4 \pi c_{s}^{2} T^{2})^{\frac{d}{2}}\Gamma \big(\frac{d}{2}\big)}\zeta(1+d)
\end{equation}
where the $\eta$ sum stands for the two modes in \nref{soundpole}, and we used the large time limit in the second equality. 
The result is remarkably independent of $D$ in this limit, and decays as a power law similarly to the diffusive one. For $d$ odd one finds instead 
\begin{equation}
\log \text{Tr}\, [e^{-i\tilde{K}T}]=V \frac{2(-1)^{\frac{d-1}{2}}\Gamma(\frac{d+1}{2})c_{s}^{d}}{(4 \pi D^{2})^{\frac{d}{2}}\Gamma(d)}\sqrt{\frac{D}{c_{s}^{2}T}}e^{-\frac{c_{s}^{2}T}{4 D}}
\end{equation}
which, unlike the result for even dimensions, is exponentially decaying and therefore much less relevant at large timescales. These two results agrees with the hydrodynamic computation in \cite{Winer:2022gqz}, which is again to be expected from holography.



\subsection{The modified boost for the $AdS_{2}$ case}\label{sec:ads2}




In this subsection we consider a black hole with an $AdS_2$ near horizon geometry. This case was already discussed in \cite{Saad:2018bqo}. Here we will review that discussion while emphasizing a few  features. 

$AdS_2$ has an $SL(2)$ isometry group  generated by $K$, $H$ and $P$ satisfying the commutation relations 
\be \la{SL2}
[K,H]=i P,~~~~~~~~~~~ [K,P]=i H~,~~~~~~~~~~~  [H,P]=i K ~.
\ee

\subsubsection{The modified boost obtained by conjugation }
 
 In this case, one can define a modified boost generator as  \cite{Saad:2018bqo}
\be \la{Smal}
\tilde{K} \propto K-i \epsilon H
\ee 
where we used $\propto$ because the modified boost can be better defined as $K$ conjugated
by an imaginary amount of $P$ evolution 
\begin{equation} \la{ModBoo}
 \tilde{K}=e^{\alpha P}K e^{-\alpha P}=\cos \alpha \, K-i \sin \alpha \, H
\end{equation}
which for small $\alpha$ reduces to \nref{Smal}. This definition is more useful because for $\alpha>0$ it makes clear that the eigenvalues of $\tilde{K}$ are independent of $\alpha$. Also, close to the horizon $P$ acts as a translation operator, so conjugating by $e^{-\alpha P}$ is the same as translating $\rho$ by $-i \alpha$ close to the horizon, which is the $i\epsilon$ prescription proposed in $\cite{Saad:2018bqo}$, which we reviewed in sec. \ref{sec:complex}.

The original Rindler patch of $AdS_{2}$ is parameterized by a radial coordinate $\rho$ and a boost time parameter $t$, so it has line element
\begin{equation} \la{AdS2Me}
	ds^{2}=-\sinh^{2}\rho \, dt^{2}+d\rho^{2}
\end{equation}
with $\rho$ going from $-\infty$ to $\infty$. 
%
%
%
%
%
For $\alpha \ll 1$, the effect of conjugating $K$ by $e^{-\alpha P}$ reproduces the general discussion that we reviewed in sec. \ref{sec:complex}, where the metric of the deformed spacetime can be written as
\begin{equation}\label{adsalpha}
	ds^2 = - \sinh^2 (\tilde{\rho} - i\alpha) dt^2+ d\tilde{\rho}^2
\end{equation}
where $\tilde{\rho}$ takes real value. The modified boost $\tilde{K}$ is the time evolution operator in this complex metric.

 \begin{figure}[h]
    \begin{center}
   \includegraphics[scale=.25]{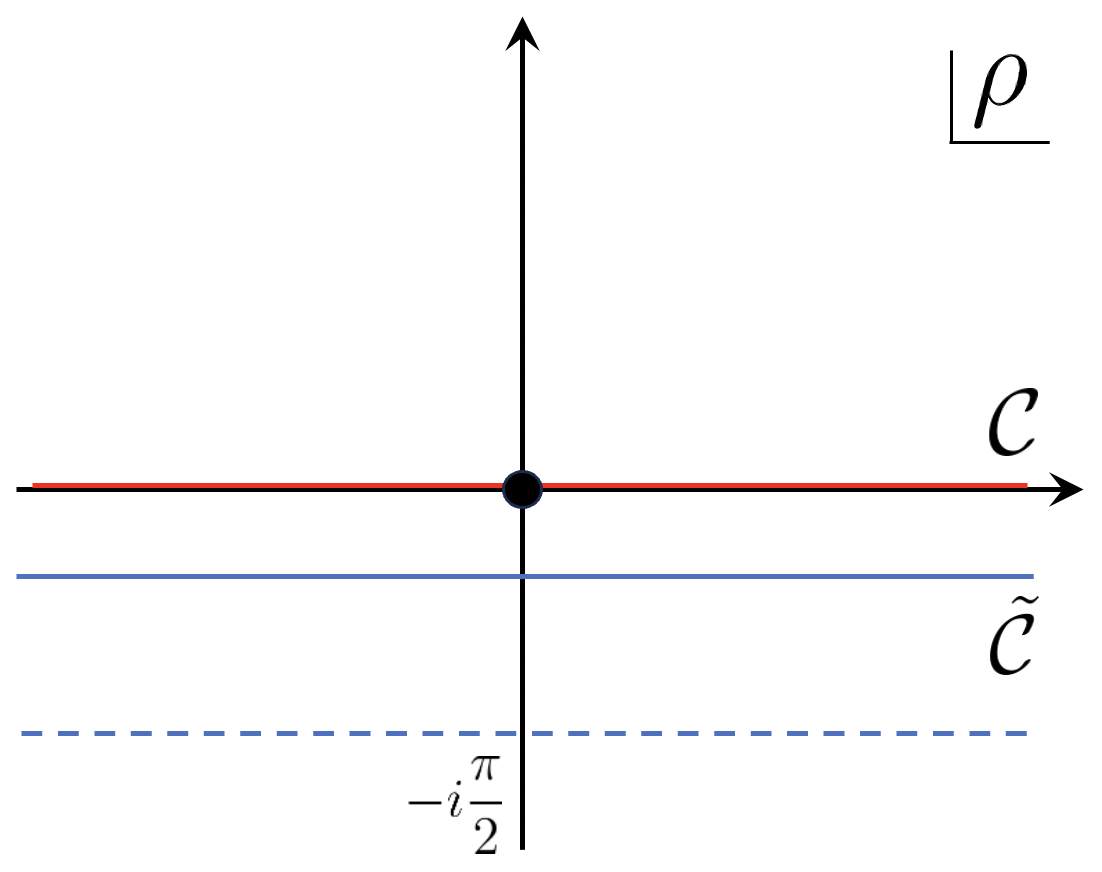}
    \end{center}
    \caption{By changing $K$ into $\tilde{K}$, we are deforming the contour away from the real $\rho$ axis $\mathcal{C}$, into a contour $\tilde{\mathcal{C}}$ with constant and negative imaginary part. Furthermore, when the imaginary part becomes $-i\pi/2$, the metric on the contour becomes that of Euclidean global $AdS_2$.   }
    \label{fig:ads2}
    \end{figure}

One special aspect of $AdS_2$ is that we can go beyond small values of $\alpha$ in (\ref{adsalpha}). In particular, there is a special value, $\alpha = \pi/2$ (see the blue dashed line in fig. \ref{fig:ads2}), where the metric becomes
\begin{equation}\label{alphapi2}
	ds^2 = \cosh^2 \tilde{\rho} \,dt^2+ d\tilde{\rho}^2~,~~~~~~~~~{\rm for } ~~\alpha = {\pi \over 2 } 
	\end{equation}
This is the metric of Euclidean $AdS_2$ and $t$ becomes the Euclidean time, the modified boost operator becomes $-i$ times the global $AdS_2$ Hamiltonian $H$, $\tilde K = - i H $. The Euclidean evolution gives an intuitive explanation in this case why the matter contribution goes to one at late time in this case.



\subsubsection{The quasinormal modes by explicit computation } 
\la{QNMExpl}

We now turn to a discussion of the quasinormal modes. For that it is convenient to define a new variable $r$ via 
%
%
\begin{equation} \la{rDefi}
     \sinh(\tilde{\rho}-i \alpha)=\frac{2 r}{1-r^{2}} ~,~~~~~~~~{\rm or }~~~~~~~~~~~r =\tanh\bigg(\frac{\tilde{\rho}-i \alpha}{2}\bigg)
\end{equation}
so that the metric becomes 
%
\begin{equation}\label{metricr}
   ds^{2}= \frac{4}{(1-r^{2})^{2}}(-r^{2}dt^{2}+dr^{2})
\end{equation}
with the boundaries at 
$r=\pm 1$. It is particularly easy to solve the wave equation for a massless scalar field with Dirichlet boundary conditions. 
Since the metric is conformally flat we just have to solve the equation of motion in flat space. The eigenstates of $\tilde{K}$   have form $\phi = e^{ - i \omega t } \phi_\omega(r)$, with $\phi_{\omega}=a_{\omega}r^{-i \omega}+b_{\omega}r^{i \omega}$.  Imposing the boundary conditions as well as continuity from $\tilde \rho$ positive to negative, which implies analyticity in $r$ in the Im$(r)< 0$ region \nref{rDefi}, we find $\omega_{n}=-i n$ with $n$ an integer different from zero. The associated solutions are 
\begin{equation}
\phi_{n}\sim(r^{-n}-r^{n}).
\end{equation}





\subsubsection{The quasinormal modes from a more elegant computation } \label{sec:quasiele}

There is another more elegant and direct way to get the quasinormal modes for any field  based on the following observation. As we observed in (\ref{alphapi2}), for $\alpha = \pi/2$ we have $\tilde K = - i H$ \nref{ModBoo}. $H$ is the generator of global time translations in $AdS_2$ and the eigenvalues of $H$ are given by $\Delta + n$, where $\Delta$ is the usual dimension of the field and $n\geq 0$. The case of the scalar field discussed above is $\Delta =1$. 
We can leverage this observation to find the eigenvalues and eigenvectors  for all other values of $\alpha $ by noticing that  
%
%
\begin{equation} \la{ConjH}
\tilde{K}=e^{\alpha P}K e^{-\alpha P}=-i e^{-(\frac{\pi}{2}-\alpha)P}H e^{(\frac{\pi}{2}-\alpha)P}
\end{equation}

The fact that we get $\tilde K$ by conjugation from $-i H$ \nref{ConjH} explains why the eigenvalues of $\tilde K$ are independent of $\alpha$. In fact, the eigenstates of 
  of $\tilde{K}$ can also be simply obtained from those of $H$. 
A non-Hermitian operator has different right and left eigenstates, with the same eigenvalue. Here they are given by 
\begin{equation}
\begin{gathered}\la{excdef}
	\ket{\{\tilde{n}_{r}\}}=e^{-(\frac{\pi}{2}-\alpha )P}\ket{\{n\}}, \quad \quad
	\bra{\{\tilde{n}_{l}\}}=\bra{\{n\}}e^{(\frac{\pi}{2}-\alpha )P},
\end{gathered}
\end{equation}
where $\{n\}=\{n_{1},n_{2},...\}$ standing for an occupation number list of the various one-particle states, and $|\{ n\}\rangle$ denoting the eigenstates of $H$.  
In particular, empty $AdS_2$ is annihilated by $\tilde K$ since it is invariant under $SL(2)$ transformations, 
 $e^{-(\frac{\pi}{2}-\alpha)P}\ket{0_{H}}=\ket{0_{H}}$, and it is the usual $AdS_2$ vacuum state. 
 The left and right eigenvectors are not orthogonal within each group. However, the left/right eigenvectors are orthogonal  to the corresponding right/left eigenvectors
 \be 
	\bra{\{\tilde{n}'_{l}\}}\ket{\{\tilde{n}_{r}\}} = \delta_{\{n'\} , \{ n  \}}.
	\ee 
Furthermore, these eigenvectors obey a kind of completeness relation that follows by conjugation from that of the eigenvectors of $H$  
%
%
\begin{equation}\label{compads2}
\sum_{\{ n\}}\ket{\{\tilde{n}_{r}\}}\bra{\{\tilde{n}_{l}\}}=e^{-(\frac{\pi}{2}-\alpha )P}\sum_{n}\ket{\{n\}}\bra{\{n\}} e^{(\frac{\pi}{2}-\alpha )P}= e^{-(\frac{\pi}{2}-\alpha)P}\,\mathbb{I}\, e^{(\frac{\pi}{2}-\alpha)P}= \mathbb{I}
\end{equation}

In conclusion,  this method    makes it clear that the eigenvalues of $\tilde{K}$ are purely negative imaginary, independent of $\alpha$, and given by the eigenvalues of $- i H$. In addition, we can also get the explicit form of the eigenvectors, or quasinormal modes.  

Note that $K$ and $\tilde K$ in \nref{ConjH} would be unitarily equivalent if $\alpha $ were purely imaginary. On the other hand, when $\alpha$ is real, the relation between the two is more subtle since $P$ is an unbounded operator so that $e^{\alpha P } |\psi\rangle $ might not be in the Hilbert space even though $|\psi \rangle $ is in it. We will discuss this in more detail in a toy model in section \ref{sec:toy}. 

We expect that quasinormal modes are useful for describing the late time behavior of field configurations. However, they cannot be used to describe some early time cases, see appendix \ref{app:incomplete}.

 \subsection{Effect of different couplings in the spectral form factor}
\label{sec:DiffCou}

In this subsection, we discuss an application of the modified boost operator $\tilde K$ for a problem that is a small modification of the double cone which was originally discussed 
  in \cite{Cotler:2022rud}. They considered 
 the analogue of the spectral form factor computed for hamiltonians with slightly different couplings, $H_L = H + g O $, $H_R = H - g O$, where one 
 expects \cite{Guhr:1997ve,simons1993universalities}
   \be \la{ExpectDi}
  \overline{ \textrm{Tr}_{\rm bdy}[ e^{ i T H_L}] \textrm{Tr}_{\rm bdy} [ e^{ - i T H_R} ]}  \propto { T \over 2 \pi } e^{ - C g^2 T} 
   \ee  
 where $C$ is some constant related to the two point functions of the operator $O$. 
 
 It turns out that the coefficient in the exponent is related to the value of $\tilde K$ in the bulk field configuration that results on the complexified geometry after turning on the boundary values of the fields associated to the coupling $g$.  
  In the rest of the section we discuss this more explicitly. 
 
While the result is general, let us give as an explicit example the case of a massless field in $AdS_2$. We choose the metric as written in (\ref{metricr})
\begin{equation}
	ds^{2}=\frac{4}{(1-r^{2})^{2}}(-r^{2}dt^{2}+dr^{2})\,.
\end{equation}
The right and left boundaries are located at $r=\pm 1$ respectively, and $r$ goes over a curve in the lower half plane connecting these two points. The precise shape of this complex curve  is not important.
Since the metric is conformally flat,   the  general solution for a massless field that is time independent is
\begin{equation}
	\chi(r)=a+ b \log r.
\end{equation}
Imposing that $\chi(\pm 1) = \pm g$ and that the field is continuous along the complex curve in the lower $r$ plane we get 
\begin{equation}
\chi=g \bigg(1-\frac{2 i}{\pi}\log r\bigg).
\end{equation}
Using this configuration we can compute the value of $\tilde{K}$ as
\begin{equation}
\tilde{K}=\frac{1}{2}\int r dr (\partial_{r}\chi)^{2}=-\frac{2 g^2}{\pi^{2}}\int_{-1}^{1} \frac{dr}{r}=-i \frac{ 2g^{2}}{ \pi}
\end{equation}
where we used a contour in the lower half plane to do the integral. 
This means that in the bulk there is a ``vacuum energy'' contribution 
\be 
\textrm{Tr}[ e^{ - i \tilde K T } ] = e^{ - { 2 \over \pi } g^2 T }
\ee 
which is of the form \nref{ExpectDi} with $C=2/\pi$. 
Of course, this also agrees with the expression discussed  in \cite{Cotler:2022rud} with 
\be 
C = -\half G_{r r}(\omega=0)
\ee 
were $G_{r r}$ is the symmetrized two point function in the boundary defined by $G_{r r}(t)=-\frac{1}{2}\big<\{\mathcal{O}(t)\, ,\mathcal{O}(0)\}\big>$. Note that in this calculation we have taken the inverse temperature of the black hole to be $2\pi$.

\section{Connection to the large $N$ limit and von Neumann algebras}\label{sec:algebra}

\subsection{The standard wormhole Hilbert space and its trace}\label{sec:algebratrace}

In this subsection,  we would like to discuss the connection between the trace we have for the double cone in \nref{WhTr} and the more standard trace we would define for the standard wormhole.  Here by standard wormhole we simply mean the spatial wormhole in the Lorentzian two sided black hole geometry.  In short, the difference is that in the first we compactify the $T_{\rm rel}$ coordinate and we do not in the second.

Let us first consider the usual wormhole, with no identifications. 

The Hilbert space for this wormhole is generated by all the quantum fields in the bulk as well as the $M$ and $T_{\rm rel}$ variables discussed around \nref{TimeSh}. More precisely, it is more convenient to introduce the variable $m = M-M_0$ where $M_0$ is some large classical mass of the background and $m$ is some deviation from it which is of order one in the $G_N \to 0$ expansion.  This is a type I$_\infty$ algebra.\footnote{See \cite{Sorce:2023fdx} for a review of various types of von Neumann algebras.} The full Hilbert space is ${\cal H}_{\rm QFT} \times {\cal L}_2(R) $, where the last factor includes $m$ and $T_{\rm rel}$ as conjugate variables. 

With these definitions, we can view this algebra as the same algebra that arises when we take the microcanonical thermofield double discussed in \cite{Chandrasekaran:2022eqq} and consider the algebra of large $N$ operators acting on both sides. In that description the $m$ and $T_{\rm rel}$ variables arise by multiplying the microcanonical thermofield double  state by functions of $m$ and evolving the right side relative to the left side.   

As a side remark, we can think of $T_{\rm rel}$ as arising from the fact that the thermofield double is breaking the time translation symmetry (though it preserves the time translation generated by $H_R- H_L$).  This breaking is very sharp in the large $N$ limit and for the canonical thermofield double state. For this reason we can call $T_{\rm rel}$ a  ``time superfluid mode'', since it is arising from the breaking of a time translation symmetry, in the same way that an ordinary superfluid mode arises from the breaking of a $U(1)$ symmetry.

%
%
In  \cite{Witten:2021unn,Chandrasekaran:2022eqq} it was argued that this  algebra can be factorized into two type II$_\infty$ factors. It was shown that that one can define finite operators 
\be 
h_R =H_R-M_0 = m + K/2 ~,~~~~~~~h_L =H_L - M_0  = m - K/2 ~,~~~~~~~ K = h_R - h_L .
\ee 
Note $K$ is a well defined operator in ${\cal H}_{QFT}$. 


The spectral form factor in the boundary theory can be viewed as the trace 
\be \la{TraTwo}
 |Z_f(iT)|^2 = \textrm{Tr}_{\rm bdy} \left[    f(H_{R,b}-M_0) e^{ - i T H_{R,b} } f(H_{L,b}-M_0) e^{  i T H_{L,b} }\right] 
\ee 
where $f$ is the window function introduced in \nref{Zf}. For the discussion in this section, we choose it to have a width that is order one.  This involves the product to two operators, one acting on each copy. In addition, the trace itself can be written as a product of two traces over each of the copies.

One is therefore tempted to think that \nref{TraTwo} might have some connection with 
\be \la{TraWhc}
Z_{\rm wh}\equiv \textrm{Tr}_{\textrm{type I}}[   f(h_R) e^{ - i  h_R T }  f(h_L) e^{ ih_L  T }]
\ee 
where this trace is taken over the standard wormhole Hilbert space. This looks particularly nice because we have two factorized operators. However, this expression is divergent as we explain below. 



After choosing the gaussian function $f(h) = \frac{1}{\sqrt{2\pi} \Delta_E}e^{-h^2/(2\Delta_E^2) }$ we find that the trace in the right hand side of \nref{TraWhc} can be written as 
\bea 
	Z_{\textrm{wh}} 
	&=& \frac{1}{2\pi \Delta_E^2}\textrm{Tr}_{\textrm{type I}}\left[e^{- \frac{m^2}{\Delta_E^2}} e^{-i KT}   e^{- \frac{K^2}{4\Delta_E^2}}\right]= \frac{1}{2\pi \Delta_E^2}\textrm{Tr}_{_{{\cal L}_2(R)} } \left[e^{- \frac{m^2}{\Delta_E^2}} \right] \textrm{Tr}_{_{{\cal H}_{\rm QFT}}}\left[  e^{-i KT}   e^{- \frac{K^2}{4\Delta_E^2}}\right]~~~~~~~~~ \la{FinTrF}
\eea
where we factorized the trace into the two Hilbert space factors discussed above. 
Now, each term has an infinity. The first term is infinite because of the infinite volume of the $T_{\rm rel}$ coordinate. The second term is infinite due to the continuous spectrum of the operator $K$. One would naively say that both infinities are of order $e^{S(M_0)}$. 
These infinities are physical. They are related to the fact that in \nref{TraTwo} there is a disconnected contribution that is formally very large, of the order of $e^{ 2 S(M_0)} $.\footnote{By disconnected contribution, we mean the contribution to (\ref{TraTwo}) coming from two disconnected geometries where the radial and time dimensions are disks,  topologically. Note that we are considering $T$ of order one, so the disconnected disk topology still gives an exponentially large answer.} 
We can think of the divergence we encountered here as a  contribution to the disconnected contribution.  

As a related point, one might also be tempted to say that \nref{TraWhc} is the product of two traces in the type II algebras discussed in \cite{Witten:2021unn,Chandrasekaran:2022eqq}. 
However, this is not the case. In fact  
\be \la{CandNo}
Z_{\rm wh } 
\not = \textrm{Tr}_{\textrm{type II}} \left[ f(h_R) e^{ -i  h_R T} \right]\textrm{Tr}_{\textrm{type II}} \left[ f(h_L) e^{ i h_L T} \right] .
\ee 
First note that for a gaussian $f$, each of the type II traces is well defined. 
We do not expect to be able to set the proportionality constant due to the ambiguity in defining the trace, and that proportionality constant is of order $e^{2 S(M_0)}$. On the other hand,  $Z_{\rm wh}$ is infinite, as we discussed previously, so that this formula does not make sense. 
However, the right hand side of \nref{CandNo} is indeed proportional (up to a $e^{ 2 S(M_0) } $ factor)  to an expected contribution to 
  \nref{TraTwo} which consists of two separate disks, the disconnected contribution. 
 
Returning to the expression in \nref{FinTrF}, 
let us   discuss first  the second factor in \nref{FinTrF}, which can be regulated by introducing the modified boost operator $\tilde{K}$. 
We will discuss a general algebraic construction of $\tilde{K}$ in sec. \ref{sec:algmod}, but for now, let us assume such an operator exists, or we simply adopt the geometric definition discussed in sec. \ref{sec:complex}. One simple way of regulating the calculation will be to replace all the $K$'s in (\ref{FinTrF}) by $\tilde{K}$. However, since $\tilde K$ has complex eigenvalues,   we  can run into trouble from a  $\tilde K^2$ term   in  \nref{FinTrF}. 
 We can imagine restricting the sum over quasinormal modes so that the $\tilde K^2$  term does not become large. 
 We can write 
\be 
\textrm{Tr}[ e^{ - i K T } e^{ - { K^2 \over 4 \Delta_E^2} } ] \propto \int d \sigma e^{ - \Delta_E^2 \sigma^2 } \textrm{Tr}[ e^{ - i (T + \sigma ) K } ] 
\ee 
So we see that the trace of the last expression is basically what we had considered in previous sections, and we can regularize it by taking $K \to \tilde K$.  This means that this trace can be viewed as putting the bulk quantum fields on the modified double cone. We will discuss the gravitational degrees of freedom below. 
  
Another way of regulating the trace in (\ref{FinTrF}) would be to only modify the boost operator in the evolution $e^{-iKT}$ to $\tilde{K}$. In other words, we are keeping the energy window function $f(h)$ fixed, but only deform  the evolution operator. This means that we need to compute $\textrm{Tr}[ e^{ - i \tilde K T} e^{ - {K^2 \over 4 \Delta_E^2 }  } ] $ and this is well defined.
\footnote{We can evaluate it in the basis of eigenstates of the ordinary boost, so that we get $\int_{-\infty}^\infty d\omega \mu(\omega) \bra{\omega} e^{-i\tilde{K}T}\ket{\omega} e^{-\omega^2/(4\Delta_E^2)}$.  We expect this integral to be well defined since $\int_{-\infty}^\infty d \omega \mu(\omega) \bra{\omega} e^{-i\tilde{K} T}\ket{\omega}$ is finite.} 
 
Of course, the trace  $ \textrm{Tr} [e^{ - i \tilde K T}] $ is computed by the path integral of  the bulk quantum fields  in the modified wormhole. This already looks a bit like the double cone. However, there is another issue that we will turn to. 

Let us now concentrate on the first factor in \nref{FinTrF}. This trace has the form 
\be 
\frac{1}{2\pi \Delta_E^2}\textrm{Tr}_{_{{\cal L}_2(R)} } \left[e^{- \frac{m^2}{\Delta_E^2}} \right] = { 1 \over 2 \pi } \int_{-\infty}^\infty  d T_{\rm rel} \int dm f(m)^2 
\ee
where we used a gaussian form for $f$. The first integral is infinite due to the volume of $T_{\rm rel} $.  

In our discussion of the double cone we have compactified  $T_{\rm rel} \sim T_{\rm rel} + T $. So we see that the trace we defined in the double cone is actually not the same as the naive trace we would define in the wormhole Hilbert space. In other words, the double cone $Z_{\textrm{dc}}$ is {\it not} equal to $Z_{\textrm{wh}}$. As we said, $Z_{\textrm{wh}}$ is infinite due to the $T_{\rm rel}$ integral while the double cone is finite. 

In this discussion  we are emphasizing the fact that even though we motivated our discussion of the measure factor from the Hilbert space of the wormhole, the compactification of $T_{\rm rel}$ is part of the definition of the Hilbert space of the double cone. We are saying that, by definition, it has $T_{\rm rel}$ compact with period $T$. In other words, the compactification of the Schwarschild time $t \sim t + T$ comes automatically because we consider ${\rm Tr}[ e^{ - i \tilde K T} ] $. But the compactification of $T_{\rm rel}$ needs to be put in ``by hand'', as part of the definition of the double cone geometry.


 In order to further understand the relation of the relation between the double cone and the ordinary wormhole, let us change the operator we consider in such a way that we get a finite trace even for the standard wormhole. Instead of \nref{FinTrF} we consider the operator 
 \be \la{HWFinT}
 \hat Z_{\rm wh} = \textrm{Tr}_{{\cal L}_2(R)} \left[ f(m)^2 g(T_{\rm rel} ) \right] \textrm{Tr}_{{\cal H}_{\rm  QFT} } \left[ e^{ - i \tilde K T} e^{ -  K^2 \over 4 \Delta_E^2 } \right] 
\ee 
where $g$ is a function that decreases sufficiently fast to have a finite integral. Now the first factor  leads to 
\be 
{ 1 \over 2 \pi } \int dm f(m)^2 \int_{-\infty}^\infty  d T_{\rm rel} \,g(T_{\rm rel}) 
\ee
and this last integral is finite. It can be rewritten as 
\be 
 \int_{-\infty}^\infty  d T_{\rm rel} \,g(T_{\rm rel}) = \int_0^T  d T_{\rm rel}\,g_p(T_{\rm rel}) , ~~~~~~~~~~g_p(T') = \sum_{n=-\infty}^\infty g( T' + n T ) 
\ee 
where we defined a new periodic function $g_p$ from the first function $g$. This second function is one we could consider in the double cone Hilbert space. 

So we see that inserting $g$ in the original wormhole trace is the same as inserting $g_p$ in the double cone problem. 
Then we conclude that 
\bea \la{HWFinT}
 \hat Z_{\rm wh} &=& \textrm{Tr}_{{\cal L}_2(R)} \left[ f(m)^2 g(T_{\rm rel} ) \right] \textrm{Tr}_{{\cal H}_{\rm QFT} } \left[ e^{ - i \tilde K T} e^{ -  K^2 \over 4 \Delta_E^2 } \right] =   \hat{Z}_{\rm dc},
 \cr 
  \hat{Z}_{\rm dc} &=&\textrm{Tr}_{{\cal L}_2(S^1)} \left[ f(m)^2 g_p(T_{\rm rel} ) \right] \textrm{Tr}_{{\cal H}_{\rm QFT} } \left[ e^{ - i \tilde K T} e^{ -  K^2 \over 4 \Delta_E^2 } \right]  
\eea 
where ${\cal L}_2(S^1)$ is indicating that we compactified $T_{\rm rel}$. 

So we conclude that after inserting a suitable operator $g$, the ordinary trace in the wormhole Hilbert space is the same as the trace in the double cone Hilbert space of a different operator $g_p$, a periodic version of $g$. 
We could choose any function $g$ as long as its integral is finite.   One example would be to take the minimal length geodesic between the left boundary at Schwarzschild time zero and and right boundary  at some time  $t$. Let us call its length  $\ell(t)$  and define  $g(t)= e^{ - \kappa \ell(t) } $. Then we see that $g_p$ looks a bit like a sum over geodesics that wind around the double cone, see the discussion in appendix A.2 of \cite{Yan:2023rjh}.  

The conclusion of this discussion is that the double cone Hilbert space is different than the usual wormhole Hilbert space. The difference is the compactification of the time $T_{\rm rel}$.

\subsection{Algebraic construction of the modified boost}\label{sec:algmod}

Here we give an alternative definition of the operator  $\tilde K$ for general black holes that highlights some its features. 

As discussed in \cite{Saad:2018bqo}, and reviewed in \ref{sec:ads2}, for the $AdS_2$ case we can 
view the operator $\tilde K$ as related by conjugation to $K$ by a complex ``translation'' 
\be 
\tilde K = e^{ \epsilon P } K e^{- \epsilon P }. 
\ee 
Near the horizon $P$ acts as a spatial translation.

 For a general black hole, we had defined $\tilde K$ as the operator generating time translations using a complexified metric following \nref{BHme} and \nref{AnalC}. This complexification of the metric is really important only near the horizon. But near the horizon we can view the metric as the metric of flat space. In that flat space approximation, besides $K$, we also have the space and time translation generators $P$ and $H$. And the complexification of the metric can indeed be viewed as a conjugation 
\be \la{KtTr} 
 \tilde K = e^{\epsilon P } K e^{ - \epsilon P } .
 \ee 
This can be understood more explicitly by looking at the shift of $\rho$ in the near horizon region given by \nref{NHmec}. If we define the flat space coordinates 
  $X^\pm = \pm \rho e^{\pm t }$, then we see that $X^1 = \half (X^+ - X^-) = \rho \cosh t $. This implies that a shift of $\rho$ downwards in the complex plane is essentially a translation in the $X^1$ direction by an imaginary amount, leading to \nref{KtTr}.
\begin{figure}[h]
    \begin{center}
    \includegraphics[scale=.3]{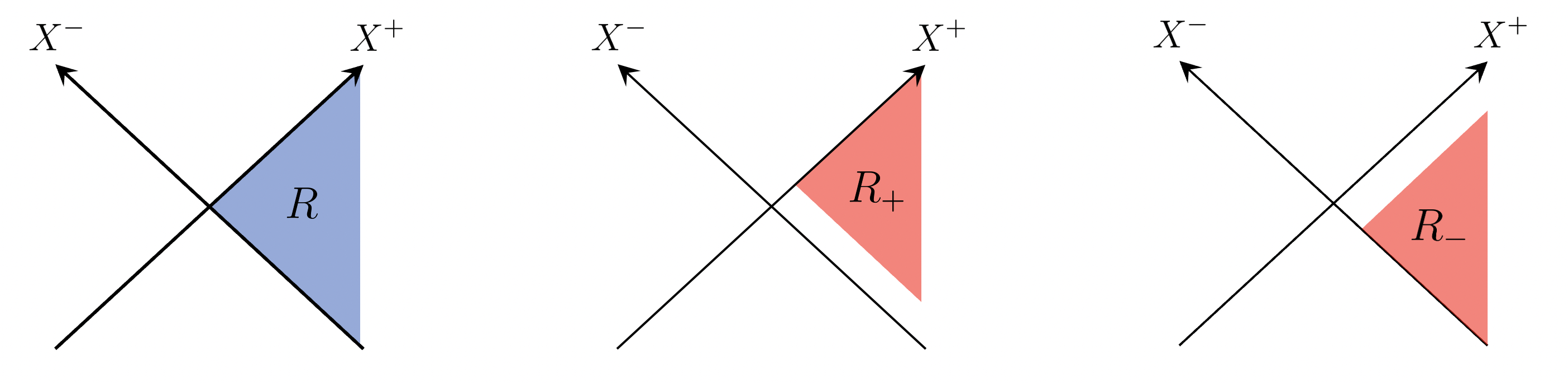}
    \end{center}
    \caption{Definition of the subregions $R$, $R_+$, $R_-$.  }
   \label{Wedges}
 \end{figure}
 
 Now, in a general black hole, $P$ and $H$ are not exact symmetries, they are only approximate symmetries in the near horizon region. However, using  features of the bulk  von Neumann algebras one can define operators we will call $\hat P$ and $\hat H$. 
They can be defined as follows in the bulk, in the quantum field theory approximation. We consider the right wedge $R$ and two subwedges $R_+$ and $R_-$ as in figure \ref{Wedges}. We then consider the modular Hamiltonians associated to these subregions. $K_R$ is what we have been calling $K$ so far. We also have $K_{R_\pm}$. 
We then define 
\be 
 \hat P^- = K_R - K_{R_+} ,~~~~~~~~~~~  \hat P^+=     K_R - K_{R_-}  .
\ee 
Using results from \cite{Borchers:1991xk,Wiesbrock:1992mg,Borchers:1998,Borchers:1998ye}  
one can see that 
the operators $\hat P^\pm$ are positive definite and annihilate the vacuum. In general they do not act in a geometrically simple way, see \cite{Leutheusser:2021frk} for some explicit formulas for free bulk fields. However, near the horizon they act as ordinary null translations. 
In addition, they can be shown to obey \cite{Borchers:1991xk,Wiesbrock:1992mg,Borchers:1998,Borchers:1998ye} 
\be \la{LorTr}
[ K , \hat P^\pm ] = \pm i \hat P^\pm .
\ee 
Note that we are not making any statements about $[\hat P^+ , \hat P^-]$, so that we should not confuse $\hat P^\pm$ with ordinary translations (or $SL(2)$ generators). They only look like  translations in the near the horizon region.\footnote{For the special case of JT gravity one can write a more explicit form for these generators  \cite{Lin:2019qwu}, where they actually act geometrically. See also a chaos based discussion of similar generators in \cite{DeBoer:2019kdj}, as well as \cite{Jefferson:2018ksk} for relation to the interior. }
  
We then define 
\begin{equation}
	\hat P= \hat P^+- \hat{P}^{-} \text{, } ~~~~~~~~~~ \hat H=   \hat P^{+}  +\hat P^{-}
\end{equation}
where we have not been carefull about the overall normalization since it depends on the definition of the regions. 
Since $\hat P^\pm$ annihilate the empty wormhole state, $\hat P$ and $\hat H$ also annihilate it. In addition, since $\hat P^\pm$ are non negative   so is $\hat H$. 



This means that we can define the operator 
\be \la{TisFC}
\tilde K =  e^{\epsilon \hat P } K e^{ - \epsilon \hat P }   .
\ee 
This formulation has some advantages which are the following. First we see that since $\hat P$ annihilates the thermofield double, then so does the operator $\tilde K$. This means that the thermofield double is a state left invariant by both $K$ and $\tilde{K}$. 
Second, the right eigenvalues of $\tilde K$ are expected to be independent of $\epsilon$. Given the right eigenstates of $\tilde K_{\epsilon}$ we can find the ones for $\tilde K_{\epsilon'}$ by acting with $e^{ ( \epsilon' - \epsilon ) \hat P}$ on the eigenstates of $\tilde K_{\epsilon}$ and they will have the same eigenvalue. 

Finally,  expanding $\tilde{K}$ to first order in $\epsilon$ we find 
\begin{equation}
	\tilde{K}=K+\epsilon[\hat P^{+}- \hat P^{-},K]+O(\epsilon^{2})=K-i \epsilon \hat H+O(\epsilon^{2})
\end{equation}
where we used \nref{LorTr}. 
%
Given that $K$ is hermitian and $\hat H$ is semipositive definite one can show that the eigenvalues of $\tilde K$ have nonpositive imaginary part.
This is proved as follows. Take, $|\lambda\rangle $  a right eigenvector of $\tilde K$, then $ \lambda = \langle \lambda| \tilde K |\lambda \rangle = \langle \lambda |K |\lambda \rangle- i \epsilon \langle \lambda | \hat H |\lambda \rangle$. Since $K$ is hermitian, the first term is real and since $\hat H \geq 0$, we find that Im$(\lambda)\leq 0$. 

Note that the deformation we propose in \nref{TisFC} is slightly different than the $i\epsilon$ prescription in \cite{Saad:2018bqo}, which is based on considering the complexified geometry. In particular, the operator in \nref{TisFC} is not expected to act geometrically, even after we complexify the geometry. However, near the horizon we expect that the two deformations act in the same way. 


In this construction,  we have ignored the time shift zero mode and we have used the bulk field theory. In particular, we have used the type III$_1$ structure of bulk quantum field theory.

\subsection{Implications for the interpretation of the double cone}


Everything discussed in the last subsection involves just the bulk theory.  However, the main point of \cite{Leutheusser:2021frk} was to offer a boundary interpretation of these operators as arising from the large $N$ limit. That implied, in particular, that their boundary versions $\hat P_b^\pm$ and also $\hat H_b$ and $\hat P_b$ are {\it two sided operators}. 
Therefore, $\tilde K_b$ defined as $\tilde K_b = e^{ \epsilon \hat P_b} K_b e^{-\epsilon \hat P_b} $, is a two sided operator, which is {\it not} just the difference of two one sided operators such  as $K_b \equiv H_R - H_L$. 

Therefore, if we think of the $i\epsilon $ prescription of Saad, Shenker and Stanford \cite{Saad:2018bqo}, not as a mathematical trick, but as a change in the problem, as in \nref{TisFC}, then we understand why evolution by $\tilde K$ involves a coupling between the two sides. 

In other words, we can interpret of the double cone wormhole, with the $i\epsilon$ prescription, in the boundary theory as
\begin{equation}\label{double}
	 \textrm{Tr}[f(h_L) 
	 e^{ - i \tilde K T } f(h_R)] = \textrm{Tr}_{\textrm{bdy}} \left[f(h_L) 
	 e^{-i \tilde{K}_bT} f(h_R)\right]_{N=\infty}
\end{equation}
where the first trace involve bulk quantities while the second is a trace in the large $N$ von Neumann algebras of the boundary theory. 
Of course $\tilde K_b$ is related by conjugation by $\hat P_b$ to $K_b$. Notice that this trace in the large $N$ algebra does not factorize.


Note that when we talked about $\hat P_b$ and $\hat H_b$ above we discussed their construction in terms of a particular large $N$ limit, as in \cite{Leutheusser:2021frk}. We have not demonstrated that they can be defined at finite $N$. 
To understand the finite $N$ construction, we need a better understanding of what type of physical operation in the boundary theory would generate $\tilde{K}$ in the bulk. In systems where a boundary theory is available, such as the SYK model, we can then have a finite $N$ definition of operators which act as the SL(2) generators on the Hilbert space of the wormhole \cite{Lin:2019qwu}, see also \cite{Harlow:2021dfp}. 

This discussion suggests that the definition of $\tilde K$ involves a coupling between the two sides. Though this seems to be true in this way of introducing the deformation, there is one other way to introduce a deformation which does not involve a coupling between the two sides. This involves imaginary couplings and we will discuss it in section \ref{ImCou}. 




As a side comment, a similar $i \epsilon $ prescription arises when we discuss hydrodynamics of thermalization \cite{Herzog:2002pc,Son:2009vu,Glorioso:2018mmw}. It can probably be discussed in a similar way. 

%

\section{Analyzing the backreaction of the matter fields in the case of JT gravity plus matter}

\label{sec:backreaction}

So far, we have studied the matter fields on the double cone solution without taking into account their backreaction on the gravity solution. It is natural to ask how the backreaction would modify the double cone solution, and, in particular, what it implies for the modified boost evolution. We will study this problem in JT gravity plus matter consisting of a conformal field theory in the bulk with a large central charge $c$. This assumption is made so that we can treat the problem using classical saddle points. 
We will find the correction to the solution from  two different perspectives in sec. \ref{sec:newsad} and \ref{sec:local}.  
It will turn out that the solution has a puzzling feature on which we comment in sec. \ref{sec:physint}.

\subsection{Finding the backreacted solution using the Schwarzian description }\label{sec:newsad}

 \subsubsection{Review of the case with no matter }

 To be concrete, we consider the spectral form factor defined in (\ref{Zf}), with specific window function
\begin{equation}
	f(E') = e^{- \frac{(E'-E)^2}{2\Delta_E^2}}.
\end{equation}
The spectral form factor $\overline{|Z_f (T)|^2}$  can then be computed as
\begin{equation}\label{Y2}
	    \overline{|Z_f (T)|^2 }= \frac{\Delta_E^2}{(\sqrt{2\pi} i )^2 }\int d\beta_L\, d\beta_R\, e^{(\beta_L + \beta_R) E + \frac{1}{2}(\beta_L^2 + \beta_R^2) \Delta_E^2 } \overline{Z(\beta_L - iT) Z(\beta_R + iT)},
\end{equation} 
where we integrate $\beta_L,\beta_R$ along the imaginary direction.  

We consider the bulk metric $ds^2 = - dt^2 \sinh \rho^2 + d\rho^2$ and write the Schwarzian variable in terms of $t_L(u_L)$ and 
$t_R(u_R)$. We consider a solution with 
\be \la{tAnd}
t \sim t + b ~,~~~~~~~~~ t_L = { b \, u_L \over \beta_L - i T} ~,~~~~~~~ t_R =  { b \, u_R \over \beta_R + i T } 
\ee 
where $u_{L,R}$ are Euclidean boundary times. 
\def\hrho{{\hat \rho}}
Here $b$ is a parameter to be integrated over. In the classical limit, we can put \nref{tAnd} into the Schwarzian action, and find that we need to extremize   
\be \la{ActNM}
(\beta_L + \beta_R) E + \half (\beta_L^2 + \beta_R^2) \Delta_E^2 - { \phi_r b^2 \over 2 } \left( { 1 \over \beta_R + i  T } + { 1 \over \beta_L - i T } \right) 
\ee 
with respect to $b$, $\beta_L$ and $\beta_R$. 
This gives 
%
%
\begin{equation}\label{saddb}
	b_* = \sqrt{\frac{2E}{\phi_r}} T, ~~~~~~~ \beta_L = \beta_R =0,
\end{equation}
which implies that we have the pure lorentzian double cone geometry. 

\subsubsection{Adding the matter }

When we add matter fields, we will be deforming the geometry to $\rho = \tilde \rho - i \epsilon$ 
\begin{equation}\label{metric2}
	ds^2 = -\sinh^2 \rho dt^2 + d\rho^2 =-  \sinh^2 (\tilde{\rho} - i\epsilon) \,dt^2 + d\tilde{\rho}^2,\quad ~~~~~~~~ t\sim t+
	b
\end{equation}
so that the matter partition function is well defined and makes an extra  contribution $\log Z_m(b)$ to the effective action \nref{ActNM}. 
 Of course, as was discussed in \cite{Saad:2018bqo} and reviewed in sec. \ref{sec:ads2}, in AdS$_2$, we can rewrite it as  
\begin{equation}\label{Zmatglobal}
	Z_m= \textrm{Tr}\left[e^{- b H}\right]
\end{equation}
where $H$ is the global $AdS_2$ Hamiltonian. 
After extremization we get 
\begin{equation}\label{saddeqn}
\begin{aligned}
	& \beta_R: \quad  E +    \beta_R \Delta^2_E + \frac{\phi_r b^2}{2 (\beta_R + i T)^2}  =0
	\\
	& \beta_L: \quad  E +    \beta_L \Delta^2_E + \frac{\phi_r b^2}{2 (\beta_L - i T)^2}  =0\\
	& b : \quad \phi_r b \left( {1 \over \beta_L - i T } + { 1 \over \beta_R + i T } \right)  +   \mathcal{E}(b) = 0,
\end{aligned}
\end{equation}
where we have defined
\begin{equation}\label{energy}
\mathcal{E}(b) \equiv - \partial_{b} \log \textrm{Tr} \left[e^{-b H}\right]
\end{equation}
%
which is a positive quantity for $b>0$. 
From this, we can immediately see from the third equation in (\ref{saddeqn}) that, at least perturbatively, we will have a saddle point value of $\beta_L + \beta_R<0$. More explicitly, we can solve (\ref{saddeqn}) perturbatively to the linear order in the backreaction, and find the saddle point:
\begin{equation}\label{backed}
	\beta_L + \beta_R \approx - \frac{\mathcal{E}(b_*)T}{\sqrt{2\phi_r E}}, \quad ~~~~
	\beta_R - \beta_L \approx -i \sqrt{ 2 E \over \phi_r} { \mathcal{E}(b_*) \over \Delta_E^2 }, ~~~~ 
	%
	\quad 
	b=b_* - \mathcal{E}(b_*) \frac{4 E^2 + \Delta^2_E T^2}{4 \phi_r E \Delta^2_E},
\end{equation}
where $b_*$ was defined in (\ref{saddb}).
Notice that since $\beta_R - \beta_L$ is purely imaginary, we can define 
\be 
T' \equiv  T - i \half (\beta_R - \beta_L)  ~,~~~~~~\beta \equiv \half (\beta_R + \beta_L)
\ee 
so that, at the saddle point, we can think of the period identifications in the gravitational path integral as $\beta \pm iT'$. Since we no longer have a purely Lorentzian evolution on the boundaries, it suggests that  the two physical boundaries are in fact located in the upper half plane of $\rho$ coordinate in (\ref{metric2}).

In particular this suggests that the saddle point geometry has the form 
\be \la{ModMetE}
ds^2 = - \sinh^2(\tilde \rho  - i \alpha ) dt^2 + d\tilde \rho^2 ~,~~~~~~ \tan \alpha = { \beta \over T' } ~~~~~{\rm or } ~~~~ \alpha \approx { \beta \over T} 
\ee 
Since $\beta < 0$, \nref{backed}, we have a deformation with the ``wrong'' sign. 
The negative saddle point value for $\beta$ is reminiscent of the discussion in \cite{Cotler:2022rud}, where a negative saddle point was also found when the couplings on the two sides of the wormhole are slightly different. In fact, though the physical setup is different,  the  mathematical reason is essentially the same. The different couplings imply that on the Im$(\rho) = -\pi/2$ section we have a scalar field with a nontrivial profile that produces a positive energy $\cal E$. Once we know this, the rest of the argument is the same, including equation \nref{backed}.

\subsection{Finding the backreacted solution using a local bulk perspective}\label{sec:local}

The negative saddle point value of $\beta$ in (\ref{backed}) can also be derived from a more local bulk perspective, by finding the form of the bulk dilaton. 
We can use the fact that in JT gravity the AdS$_2$ metric is rigid and stick to a fixed metric
\begin{equation}\label{met3}
	ds^2 = -\sinh^2 \rho dt^2 + d\rho^2, \quad t\sim t+b, 
\end{equation}
and determine where the asymptotic boundaries are located in this coordinate system. This  location is set by   the physical requirement that the dilaton $\phi$ goes to positive infinity there. Without backreaction, the boundaries are simply at $\rho = \pm\infty$ as usual, but as we will explain, this changes when we take backreaction into account.

With no backreaction taken into account, the dilaton profile on (\ref{met3}) is given by
\begin{equation}
	\phi = \phi_h \cosh \rho,
\end{equation}
where $\phi_h$ is the value of dilaton at $\rho=0$, and in our conventions, it is related to $\phi_r,E$ by $\phi_h = \sqrt{2\phi_rE}$. Now, the nonzero stress tensor coming from the matter fields   generates a new piece for the dilaton, which we find as follows. The SSS prescription for (\ref{met3}) is that we should deform the contour for $\rho$ into the lower half plane. In particular, we can deform it to $\rho = -i\pi/2 + \rho'$, where in terms of $\rho'$ the metric looks like
\begin{equation}\label{met4}
	ds^2 = \cosh^2 \rho' dt^2 + d\rho'^2, \quad t\sim t+b.
\end{equation}
On the slice parametrized by real $\rho'$, the metric (\ref{met4}) is Euclidean global AdS$_2$. Since the time direction is compactified with period $b$, the matter fields are in a thermal state, therefore having a nonzero stress tensor. For simplicity we assume that the matter fields are described by a conformal field theory.\footnote{The vacuum contribution to the stress tensor cancels out the contribution from the conformal anomaly.} For convenience, we do a further coordinate transformation
\begin{equation}
	\frac{1}{\sin\sigma'} = \cosh\rho', \quad 0<\sigma'<\pi,
\end{equation}
such that the metric becomes
\begin{equation}
	ds^2 = \frac{dt^2 + d\sigma'^2}{\sin^2 \sigma'}, \quad 0<\sigma'<\pi .
\end{equation}
Since we have a conformal field theory, the stress tensor 
is the same as the one on a (compactified) flat strip
\begin{equation}\label{stress}
	T_{tt} = - T_{\sigma'\sigma'} = -\frac{\mathcal{E}}{\pi},
\end{equation}
where $\mathcal{E}$ is the total energy on the strip, which is equivalent to what we defined in (\ref{energy}). The dilaton profile that sourced by (\ref{stress}) can be found in Appendix C of \cite{Maldacena:2018lmt}, which gives\footnote{In \cite{Maldacena:2018lmt}, the energy was negative due to coupling the two sides. Here instead we have positive energy so there is a sign difference.}
\begin{equation}
	\delta \phi = - \frac{\mathcal{E}}{\pi} \left( \frac{\frac{\pi}{2} - \sigma'}{\tan \sigma'} +1\right) = -  \frac{\mathcal{E}}{\pi} \left[ \sinh \rho' \left(2 \arctan e^{\rho'} - \frac{\pi}{2}\right) + 1\right].
\end{equation}
So in terms of the original coordinate $\rho$, the total dilaton is given by
\begin{equation}\label{phitotal}
	\phi = \phi_h \cosh \rho -  2i\frac{\mathcal{E}}{\pi} \left[ \cosh \rho \arctan \left(\tanh \frac{i\frac{\pi}{2} + \rho }{2}\right) - \frac{i}{2}\right].
\end{equation}
With the expression for the dilaton, we can now look at when it approaches $+\infty$, which determines where the physical boundaries are located in coordinate $\rho$. For the right boundary, we have $\rho_b = \rho_c + i\delta \rho$, with    real $\rho_c\gg 1$, and
\begin{equation}\label{phib}
	\phi_b = \phi (\rho_c + i\delta \rho) \approx \frac{e^{\rho_c}}{2} \left( \phi_h + i \phi_h \delta \rho-i \frac{\mathcal{E}}{2}\right),
\end{equation}
where we only kept the expansion to linear order of $\delta\rho\sim \mathcal{E}$. Note that $\phi_h$ is real to leading order. Then, 
demanding (\ref{phib}) to be real we get 
\begin{equation}\label{deltarho}
	\delta \rho =  \frac{\mathcal{E}}{2 \phi_h} =  \frac{\mathcal{E}}{2 \sqrt{2 \phi_r E}}. 
\end{equation}
In principle, $\phi_h$ could get an imaginary part at linear order in $\mathcal{E}$. However, by demanding that the boundary condition is also satisfied on the left, i.e. $\rho = - \rho_c + i \delta\rho$, with the same $\delta \rho$ determined by (\ref{deltarho}), $\phi_h$ is constrained to remain real at linear order. 
 
\begin{figure}[h!]
    \begin{center}
   \includegraphics[scale=.18]{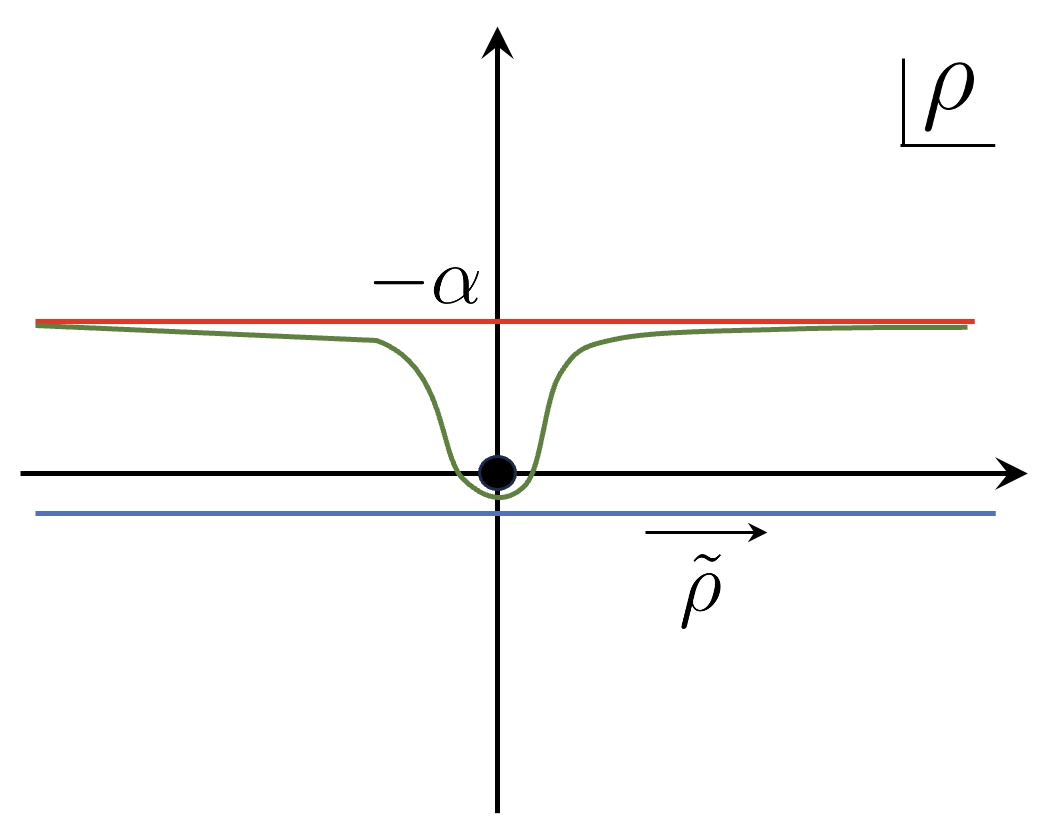}
    \end{center}
    \caption{When we take into account the backreaction of the matter field, the physical boundaries of the spacetime are no longer on the real axis of $\rho$, but instead become in the upper half plane, at the two ends of the red line.  The generalized SSS prescription would be to take a contour (shown in green) that takes a detour below $\rho=0$. Another possibility is to take a contour that is parametrized by real parameter $\tilde{\rho}$ (shown in blue).  }
    \label{fig:contour2}
 \end{figure}

We can now compare the result (\ref{deltarho}) with what we had gotten in \ref{sec:newsad}. There we found that the physical boundaries, relative to the slice where the metric takes the form of (\ref{met3}), has a positive displacement $-\alpha$ in the imaginary direction (see fig. \ref{fig:contour2}), determined by \nref{ModMetE}
and (\ref{backed})
\begin{equation}
	-\alpha = - \arctan \frac{\beta}{T'} =\frac{\mathcal{E} }{2 \sqrt{2\phi_r E}} + \mathcal{O} (\mathcal{E}^2).
\end{equation}
We see that it matches precisely with (\ref{deltarho}) at leading order in $\mathcal{E}$. It would be interesting to also understand how to compute the other quantities in (\ref{backed}) from this local perspective. Note that it would involve taking into account the energy width $\Delta_E$ properly, which we have not done in this calculation. 

It is interesting to note that 
the backreacted dilaton is singular at the bifurcating point $\rho = 0$
\begin{equation}
	\phi  \sim - \frac{\mathcal{E}}{\pi} \log \rho , \quad\quad \rho \rightarrow 0 .
\end{equation}
The singular behavior comes from the singular behavior of the stress tensor for the matter field near $\rho=0$. To see this more explicitly, we can use (\ref{stress}) and work out the stress tensor in terms of the original coordinates $\rho,t$.  One finds
\begin{equation}
	T_{tt} = - \frac{\mathcal{E}}{\pi}, \quad ~~~~~~ T_{\rho\rho} = - \frac{1}{\sinh^2 \rho} \frac{\mathcal{E}}{\pi},
\end{equation}
from which we see that $T_{\rho\rho}$ is singular at $\rho=0$. The singularity becomes more intuitive if we go to the light cone coordinates $X^\pm = \pm\tanh \frac{\rho}{2} e^{\pm t}$. We have
\begin{equation}
	T_{++} = -\frac{\mathcal{E}}{2\pi} \frac{1}{(X^+)^2}, \quad ~~~~~~~~~~ T_{--} = - \frac{\mathcal{E}}{2\pi} \frac{1}{(X^-)^2}.
\end{equation}
In other words, we find that the stress tensor is singular both at the past and future horizon. Of course, both horizons are not in the double cone geometry due to the quotient in time as well as the complex deformation. Here we are analytically continuing the stress tensor to the horizons. 

\subsection{A double cone wormhole-like solution with the right sign of displacement} 
\la{ImCou}

In this section,  we consider a modified problem which results in a deformation of the geometry which does indeed have the ``right sign''. This involves adding an imaginary coupling in such a way that the microscopic computation is 
\be 
\textrm{Tr}_{\rm bdy} \left[ e^{  i T ( H_{L,b} - ig O_{L,b}) } \right] \textrm{Tr}_{\rm bdy} \left[ e^{ - i T ( H_{R,b} + i gO_{R,b}) } \right] 
\ee 
Notice that the total evolution operator is non-hermitian and the left side operator is the hermitian conjugate of the right side one (we are assuming that $O$ is hermitian and $g$ is real).  
We can consider the simplest case where $O$ is dual to a bulk massless scalar field $\chi$. We are then putting boundary values at the left and right boundary of the form $\chi_R = i g$, $\chi_L = -i g$. This problem was considered in \cite{Garcia-Garcia:2020ttf,DouglasPrivate}.   The result of putting two different boundary values is that the scalar field will have a gradient across the bulk  and it  leads to a negative energy. The same problem but with two different real values of the scalar was studied in \cite{Cotler:2022rud} and discussed in section \ref{sec:DiffCou}, where it lead to positive energy. The sign changes due to the extra $i$.

If we imagine a large number of such fields, then we can use the classical backreacion analysis of section \ref{sec:newsad}, and formula \nref{backed} where now ${\cal E} < 0$ and then $\beta > 0$ which through \nref{ModMetE} leads to a deformation with the 
``right'' sign. 

Notice, that this would additionally lead to an extra contribution proportional to $e^{ - b_* {\cal E}(b_*)}$ which grows exponentially with $T$, after using \nref{saddb} and that ${\cal E} < 0$. We could subtract this by modifying the boundary Hamiltonian by the addition of an imaginary ground state energy. This ground state energy would be state dependent, in the sense that it would depend on the value of $E$, the energy of the background solution for the double cone (or mass of the black hole).

A natural question is the following. Since the interpretation of the double cone suggests some average over couplings in the microscopic boundary theory,  one could wonder what happens if we average over some couplings that are directly visible in the bulk theory. For example, we can consider a bulk scalar field and we could average over possible boundary conditions. One might have expected that this would induce a deformation of the geometry with the right sign of $\beta$. In appendix 
 \ref{AveAppendix}, we analyzed this problem and we found that this is not the case, at least for small conformal dimensions where the computation is best defined and  to the level of approximation that we analyzed there. 

This case with imaginary couplings can be realized in a case where the microscopic theory looks unique, by using the duality between $AdS_5 \times S^5$ and ${\cal N}=4$ super Yang Mills. We can embed a nearly  $AdS_2$ spacetime by considering the BPS black hole discussed in \cite{Gutowski:2004yv} whose near horizon description can be done using JT gravity \cite{Boruch:2022tno}. In such a solution,  the dilaton field is a constant. We can then change the boundary value of the dilaton by giving a small imaginary part to the coupling constant.

\subsection{Comments on the physical interpretation of the negative value of $\beta$ }\label{sec:physint}


We have concluded  that the inclusion of backreaction implies a complexification of the geometry with the wrong sign, see figure 
\ref{fig:contour2}. This looks like a contradiction. Namely, in order to evaluate the matter partition function we assumed a complex deformation with the right sign, however, the final geometry, after taking into account the matter backreaction,  gives a complex deformation with the opposite sign, meaning that the original computation of the matter partition function was incorrect.

We could address this problem in two ways. 

First,  one could view the choice of contour that defines the geometry as a mathematical choice and shift it to the green line in figure \ref{fig:contour2},  which would give the answer we want. This is still problematic because the contour continues to violate the  Kontsevich-Segal-Witten criterion \cite{Kontsevich:2021dmb,Witten:2021nzp}.
 More explicitly, we imagine performing the full path integral in the following order.\footnote{We thank Zhenbin Yang for suggesting this argument.} We leave the integration over $\beta $ and $b$ as the last integrals in the functional integral. First we set $\textrm{Re}(\beta ) > 0$ and we do the path integral over all the matter fields on a double cone wormhole with the right sign deformation, and a fixed $b$. This integral is well defined and gives a certain function of $\beta$ and $b$. Let us call this result $Z_m(b,\beta)$.  In principle, we are supposed to do the integral over $\beta$ along a contour along the imaginary direction along a line with $\textrm{Re}(\beta)> 0$. All along this integration contour  the matter partition function is well defined,  for positive $b$. We now imagine evaluating that integral by a saddle point evaluation. For that purpose we analytically continue $Z_m(b,\beta)$ and find the saddle point. We happen to find the saddle point at $\beta< 0$. But by using the above logic it looks like we can trust it. 
 Furthermore, the results we get using this saddle point look physically reasonable. We get a correction that will become small when $T$ becomes large (keeping $E$ fixed). It becomes small because $b_*$ will become large so that ${\cal E}(b_*) \to 0$, and $\beta \to 0$, see \nref{saddb} and \nref{backed}.

Second, we can modify the computation that we are doing in such a way that it results in a shift of contour in the imaginary direction so that the final contour is on the ``right'' side, as in the blue contour in  figure \ref{fig:contour2}. 
One possible way to generate this deformation is to add imaginary couplings, as discussed in section \ref{ImCou}.  
We could then imagine the original double cone as a limit where we remove the coupling. However, in the case with matter backreaction we would still need some small but finite value of the coupling in order to have $\beta > 0$. This sounds unreasonable since the spectral form factor  is expected, under general reasons, to give the answer of the form that we are obtaining from the original solution with $\beta < 0$. 


It is worth noting the following. 
It is useful to discuss the physical $SL(2)$ generators introduced in \cite{Lin:2019qwu}, see also \cite{Harlow:2021dfp}. The idea is that the quotient that leads to the identification $t \to t +b$ is generated by a linear combination of such generators. Clearly, this quotient is modifying the physics, therefore it should correspond to one of these physical generators. 
We can compute these generators by considering the positions of the boundary as discussed in appendix \ref{app:gaugeinv}. In the geometry that we get after including the  backreaction,   the generator that corresponds to the identification, or to $\tilde K$,  is given by 
\be \la{MOdBo}
\tilde K = G_K - i \alpha  G_H  ~, ~~~~~~ \alpha \sim {  \beta \over T}, 
\ee 
where $G_K$ and $G_H$ are the $K$ and $H$ physical SL(2) generators that act on the wormhole Hilbert space. We see that the sign of  $\alpha$ is correlated to the sign of $\beta$.  It is negative, or the wrong sign, for the solution with matter backreaction
\nref{backed}.  It is positive, or the right sign, for  the problem with imaginary couplings of section \ref{ImCou}.  

In the case with imaginary couplings, the discussion in section \ref{ImCou} suggests that 
\be 
\tilde K \sim  (H_{R,b} + i g O_{R,b}) - ( H_{L,b} - i g O_{L,b})
\ee 
with $\tilde K$ as in \nref{MOdBo}. In other words, in this case we claim that the operator that has a factorized form in the microscopic theory is $\tilde K$ rather than $K$. 
 
Therefore, once backreaction is taken into account, if we want to find a classical saddle such that the one loop partition function is well defined around that saddle, then we should introduce the imaginary couplings. Otherwise, we should imagine that we first consider an off-shell geometry such that we can evaluate the matter partition function, and then, after getting the answer for the matter path integral, we analytically continue the answer to the negative value of $\beta$. 

It should be noted that such negative values of $\beta$ have also appeared previously in \cite{Bah:2022uyz}, where they also lead to physically reasonable answers. 


\section{A Harmonic oscillator toy model}\label{sec:toy}

One surprising feature of the operators $K$ and $\tilde K$ is that they have radically different spectra. It seems surprising that a ``small'' deformation can radically change the spectrum. In this section, in order to understand better these issues, we study a toy model which contains operators analogous to $K$ and $\tilde K$. Using this toy model we will explore various issues such as the connection between the eigenvalues and eigenvectors of both operators, the behavior of matrix elements as we take $\epsilon \to 0$, the connection between the traces involving both operators, etc.

The toy model is based on the Hilbert space of $L^2$ normalizable function on a line. Using the usual $a$ and $a^\dagger$ creation and annihilation operators of a harmonic oscillators we define the operators 
\begin{equation}
\begin{gathered}
    K=\frac{i}{4}\left({a^{\dagger}}^{2}-a^{2}\right) ,\quad 
    P=-\frac{1}{4}\left(a^{2}+{a^{\dagger}}^{2}\right) ,\quad
    H=\frac{1}{2}\left(a^{\dagger}a+\frac{1}{2}\right)
\end{gathered}    
\end{equation}
which obey the SL(2) commutations relations \nref{SL2}.
We define 
\begin{equation} \la{ModBde}
    \tilde{K}_\alpha =e^{\alpha P}K e^{-\alpha P}= \cos \alpha\, K- i \sin \alpha\, H = e^{ - ( \pi/2 -\alpha )P} ( - i H) e^{ (\pi/2 - \alpha )P  } .
    \ee 
    %
%
For $\alpha=0$ we have the original $K$. For $\alpha = \pi/2$ we have $-i H$, similar to the $AdS_2$ case due to the same algebra. We call operators with $0 < \alpha \leq \pi/2$ ``modified boost'', and we sometimes drop the index $\alpha$ in $\tilde K_\alpha$ \nref{ModBde}. The idea is that for small $\alpha \sim \epsilon$ we have a deformation of a unitary operator $K$ by adding something with a negative imaginary part, namely $- i \epsilon  H $. We will be particularly interested in the $\alpha \to 0$ limit. 

\subsection{Properties of the original boost $K$} 

One can rewrite $K$ in terms of position and momentum operators  
\begin{equation}
    K=\frac{1}{4}(x p+p x)=\frac{1}{2}x p-\frac{i}{4} = - \frac{i}{2} \left[ x \partial_x + \half \right] .
\end{equation}
 This operator has real eigenvalues labelled by $\omega$, and its eigenfunctions which are delta function normalizable are \be \la{Eigfphi}
\langle x |\omega\rangle_+ =  \frac{1}{\sqrt{\pi}} x^{2 i \omega-\half  } \theta(x) ~,~~~~~~~~~~~~~\langle x | \omega \rangle_- =\frac{1}{\sqrt{\pi}}(-x)^{ 2 i \omega -\half} \theta(-x) .
\ee 
Under the evolution by $K$, a wavefunction $\psi(x)$ transforms in a simple way 
\be \la{Kev}
\langle x | e^{ - i K t } |\psi \rangle = e^{ - t/4} \psi( e^{ -t/2} x ) .
\ee 
If the wavefunction is analytic at the origin, this implies that it will decay exponentially in  time with powers as $e^{ - t ( { 1 \over 4 } + { n \over 2} ) } $. 

It is convenient to define a new variable $\phi$ via $|x| = e^{-\phi/2}$   and new wavefunctions via $\tilde \psi_\pm (\phi)  = e^{ -\phi/4} \psi( \pm e^{ - \phi/2} )$. We can now think of the Hilbert space as being given by normalizable functions describing particles living on two separate real lines. The statement is precise since the inner product of two wavefunctions can be expressed as
\begin{equation}
	\langle \chi |   \psi \rangle =\int_{-\infty}^{\infty} dx \chi^{*}(x)\psi(x)=\int_{-\infty}^{\infty} d\phi \,\tilde{\chi}_+^{*}(\phi )\tilde{\psi}_+(\phi)+
\int_{-\infty}^{\infty} d\phi \,\tilde{\chi}_-^{*}(\phi )\tilde{\psi}_-(\phi) \la{NorPhi}
\end{equation}
so that we have the usual norm in terms of the   $\tilde \psi_\pm(\phi)$ functions. 
Now $K  $ becomes just translations on these two lines, $K =  i \partial_\phi$ and the eigenvectors are just plane waves. We also that 
$\langle \phi |e^{ - i t K } |\tilde \psi_\pm \rangle = \tilde \psi_\pm ( t + \phi)$. See fig. \ref{fig:map} for an illustration. 
\begin{figure}[h]
    \begin{center}
    \includegraphics[scale=.3]{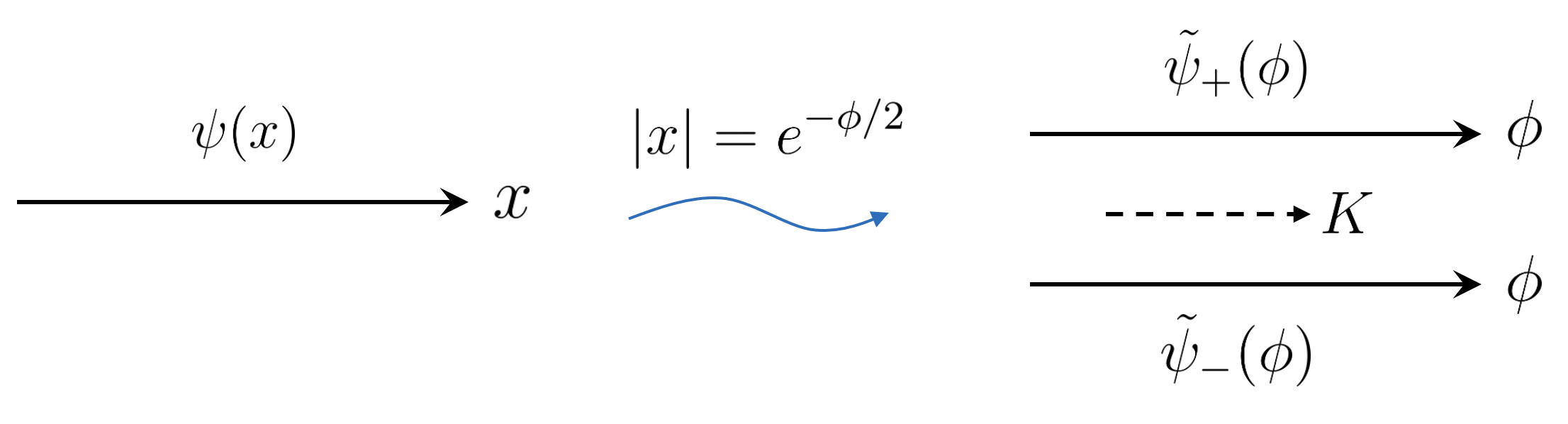}
    \end{center}
    \caption{We can map the real axis of $x$ into two real lines parameterized by $\phi$, in which $K$ acts simply as translation. 
    }
   \label{fig:map}
 \end{figure}

Another noteworthy property of $K$ is that the eigenvalue equation $K\Psi=\omega \Psi$ has formal solutions of the form $x^{2 i \omega -\half } $ for general complex $\omega $. If $\omega$ is not real, they are not normalizable.  Among these complex solutions, the following ones will play a role 
\be \la{KnEig}
\omega_n = -\frac{i}{2}\left(n+\frac{1}{2}\right)~,~~~~~~~~~~  \Psi_{n}=x^{n} \,.
\ee 
They will appear as limits of the normalizable eigenvectors of $\tilde K_\alpha$ as $\alpha \to 0$.

\subsection{Properties of the modified boost operator $\tilde K$}

We now explore the eigenvalues and eigenvectors of $\tilde K$. 
The eigenvalues and eigenstates can be constructed most directly by starting with the case $\alpha = \pi/2$ where we get $\tilde K_{\pi/2} = - i H$, in the same fashion as in sec. \ref{sec:quasiele}. This is just the usual Harmonic oscillator problem and we have the usual eigenvalues and eigenfunctions. This translates into the following eigenvalues of $\tilde K$ 
\be \la{Eigv}
 \omega_n = - {i \over 2 }  \left(  n + { 1 \over 2 } \right).
 \ee 
 The problem with general values of $\alpha$ can be obtained by conjugation as in the last formula in \nref{ModBde}. This means that the 
 eigenvalues \nref{Eigv} are the same for all $\alpha$. 
 The eigenvectors can also be obtained simply by 
\be 
\label{neigen}
|n_r\rangle =e^{ - (\frac{\pi}{2}-\alpha) P} |n \rangle ~,~~~~~~~~ \langle n_l | = \langle n| e^{(\frac{\pi}{2} - \alpha ) P }
\ee 
Since the operator $\tilde K$ is not hermitian, we have different left and right eigenvectors. 
Note that \nref{neigen} implies that left and right eigenstates have unit overlap.

 In order to write more explicit formulas it is convenient to define the operators 
%
%
\bea
\la{mdfa}
    a_{r}&=&e^{ - ( \frac{\pi}{2}-\alpha) P} a e^{   ( \frac{\pi}{2}-\alpha) P}= \cos ( {\pi \over 4} - { \alpha \over 2 })  a - 
    \sin (  {\pi \over 4 } - {\alpha \over 2} )a ^\dagger ~,~~~~~~
 \cr 
  a_{l}^\dagger &=&e^{ - ( \frac{\pi}{2}-\alpha) P} a^\dagger e^{   ( \frac{\pi}{2}-\alpha) P}= \cos ( {\pi \over 4} - { \alpha \over 2 })  a^{\dagger} + 
    \sin (  {\pi \over 4 } - {\alpha \over 2} )a  ~,~~~~~~
    \eea
%
which obey the commutations relations $[a_{r},a_{l}^{\dagger}]=[a_{l},a_{r}^{\dagger}]=1$ and $[a_{r},a_{l}]=[a_{l}^{\dagger},a_{r}^{\dagger}]=\cos \alpha$. Note that these are not standard creation and annihilation operators.  In terms of these  operators the modified boost has the simple expression 
\begin{equation}
    \tilde{K}=-\frac{i}{2} \bigg(a_{l}^{\dagger}a_{r}+\frac{1}{2}\bigg).
\end{equation}
Then, for example, the state $|0_r\rangle$ is given by $a_r |0_r \rangle =0$. Note that the analogy with \nref{sec:ads2} is not exact since, unlike the ground state in \nref{excdef}, $\ket{0_{r}}$ is not annihilated by all the generators. In fact, no state in this Hilbert space can be annihilated by $H$, since it is bounded below by $\frac{1}{4}$.\footnote{One might wonder whether this construction  is related to the single particle Hilbert spaces in AdS$_2$. It is not exactly the same since here $L_- \propto (a^\dagger)^2$ shifts the value of $L_0 \propto H$ by two instead of one in AdS$_2$. However, one can use similar constructions with harmonic oscillators to construct the correct representations corresponding to single particles in AdS$_2$, see for example \cite{Hirayama:1993ve}.  
}  Expressing $a$ and $a^\dagger $ in \nref{mdfa} in terms of $x$ and $p$, this becomes a first order differential equation that can be easily solved. In fact, this is a complexified ``squeezed'' state, similar to the ones that appear when we do Bogoliubov transformations.\footnote{The difference is that here we are conjugating by a non-unitary operator.} In this way, we   find 
\begin{equation}
	\la{lrground}
    \psi_{0,r}(x)=\bigg(\frac{1}{2 \pi \cos^{2}\frac{\alpha}{2}}\bigg)^{\frac{1}{4}}e^{-\frac{1}{2}\tan\big(\frac{\alpha}{2}\big)x^{2}},\quad ~~~~~~~~
    \psi_{0,l}(x)=\bigg(\frac{1}{2 \pi \sin^{2}\frac{\alpha}{2}}\bigg)^{\frac{1}{4}}e^{-\frac{1}{2}\cot\big(\frac{\alpha}{2}\big)x^{2}}.
\end{equation}
We    see that they are normalizable for all $ 0 < \alpha \leq \pi/2$. They become equal when $\alpha = \pi/2$, as expected. 
As $\alpha\to 0$,  they become non-normalizable. We can still formally take the limit $\alpha \to 0$.  For $\psi_{0,r}$ one gets a constant wavefunction, which is the $n=0$ case of \nref{KnEig}. On the other hand,  $\psi_{0,l}$ becomes proportional to a delta function, which is also not normalizable.    
%
 In appendix  \ref{ToyApp}, we give the explicit form for the eigenfunctions for $n>0$,  and we also show that the right eigenfunctions become simple powers as in \nref{KnEig} as $\alpha \to 0$.

We therefore see that the spectrum of $K$ and $\tilde{K}$ are different because the $\alpha \to 0$ limit takes the eigenstates of $\tilde{K}$ we described outside of the Hilbert space.   This is reminiscent of the discussion in \cite{Arean:2023ejh}, where it discussed the stability of the spectrum of non-hermitian operators in the context of quasinormal modes. This sensitivity for a given eigenvalue $\lambda_{n}$ is characterized by 
\begin{equation}
	\kappa_{n}=\frac{\sqrt{\bra{n_{r}}\ket{n_{r}}\bra{n_{l}}\ket{n_{l}}}}{|\bra{n_{l}\ket{n_{r}}}}
\end{equation}
with $\ket{n_{r}}$ and $\ket{n_{l}}$ the right and left eigenstates of the operator. The idea is that a perturbation of strength $\epsilon$ would change the eigenvalue by an amount of order $|\delta \lambda |  < \kappa_n \epsilon$ \cite{Arean:2023ejh}.    
 One can study the  spectrum stability of $\tilde{K}$ near $\alpha \sim 0$ by computing these numbers. For example,  
 \begin{equation}
	\kappa_{0}=\frac{1}{\sqrt{\sin \alpha}} \sim \frac{1}{\sqrt{\alpha}} ~,~~~~~~~{\rm for }~~~ \alpha \sim 0
\end{equation} 
which we see becomes infinitely large as $\alpha \rightarrow 0$, and explains why a small $\epsilon H$ deformation is enough to completely change the spectrum from $\tilde{K}$ to $K$.

Another important comment is that the left and right eigenvectors obey the following completeness relation 
\begin{equation} \la{CompM}
  \mathbb{I} =\sum_{n=0}^{\infty}\ket{n_{r}}\bra{n_{l}}=e^{-(\frac{\pi}{2}-\alpha)P}\left(\sum_{n=0}^{\infty}\ket{n}\bra{n}\right)e^{(\frac{\pi}{2}-\alpha)P}
\end{equation}
which follows from conjugation from the usual completeness relation for the eigenfunctions $|n\rangle$ of $H$. 
In appendix \ref{ToyApp} we prove this identity more explicitly, see  \nref{Compleness}.

\subsection{Matrix elements of the evolution operators}

Given that the eigenstates and eigenvalues of $K$ and $\tilde K$ look rather different, 
one might wonder whether simple matrix elements evolved with $e^{-i \tilde{K}t}$  agree with those of $e^{ - i  K  t}$ in the $\alpha \rightarrow 0$ limit. 
As a simple example, we can consider the matrix element of the evolution in the ground state $\ket{0}$ of $H$.
We can use the right eigenstates of $\tilde{K}$ to compute the evolution by inserting the identity \nref{CompM}
\begin{equation}
    e^{-i\tilde{K}t}\ket{0}=\sum_{n}e^{-t/4}e^{-n t/2}\bra{n_{l}}\ket{0} \ket{n_{r}}
\end{equation}
After computing the overlap of the hamiltonian ground state with $\ket{n_{l}}$ and $\ket{n_{r}}$ one can evaluate the overlap of the state above with $\ket{0}$ to find, see appendix \ref{ToyApp}, 
\begin{equation}
\begin{gathered} \la{Kanalog}
    \bra{0}e^{-i \tilde{K}t}\ket{0}=\sqrt{\frac{1}{\cosh \frac{t}{2}+ \sin \alpha \sinh \frac{t}{2}}}.
\end{gathered}
\end{equation}
In this form, we clearly see the effect of quasinormal mode decays with frequencies \nref{Eigv}. 

In particular, we can show that this agrees with the same evolution induced by the regular boost as $\alpha \rightarrow 0$. Indeed, one can explicitly compute, see appendix \ref{MatKApp},
\begin{equation}
	\label{OriKovrlp}
    \bra{0}e^{-i K t}\ket{0}= \sum_{\pm} \int d\omega\,  e^{-i\omega t}|\langle \omega_\pm | 0\rangle|^{2}=\int_{-\infty}^\infty \frac{d\omega}{ \pi \sqrt{2 \pi}}\Gamma\bigg(i \omega+\frac{1}{4}\bigg)\Gamma\bigg(-i \omega+\frac{1}{4}\bigg)e^{-i \omega t}
\end{equation}
For 
  $t>0$ we can close the contour in the lower half plane  and evaluate the integral as a sum over poles to obtain

\begin{equation}
\la{Oribostcor}	
    \bra{0}e^{-i K t}\ket{0}=
    e^{-t/4}\sqrt{\frac{2}{1+e^{-t}}}
\end{equation}
which is the same as the $\alpha \rightarrow 0$ limit of \nref{Kanalog}, so that we get a smooth limit. 
 We expect similar agreement for other matrix elements involving excited states of the original harmonic oscillator.

\subsection{Trace of the evolution operator}

A quantity where the limit $\alpha \to 0$ is more subtle is  the trace of $e^{-i K T}$. 
%
%
Since $K$ is a Hermitian operator, this will be a sum over phases that does not obviously converge. However, replacing $K \rightarrow \tilde{K}$, one obtains an obviously convergent sum. 
In fact, we can write the trace of any operator in terms of a sum over the left and right eigenstates of $\tilde K$. 
%
\begin{equation}
\text{Tr}\, A=\sum_{b}\bra{b}A\ket{b}=\sum_{b, n, m} \bra{b}\ket{n_{r}}\bra{n_{l}}A \ket{m_{r}}\bra{m_{l}}\ket{b}=\sum_{n}\bra{n_{l}}A\ket{n_{r}}
\end{equation}
where $\ket{b}$ is any complete orthonormal basis in the Hilbert space. Here we used \nref{CompM} twice, as well as 
$\sum_{b}\ket{b}\bra{b}=\mathbb{I}$ and that $\bra{m_{l}}\ket{n_{r}}=\delta_{n,m}$. 
This means that 
%
\begin{equation} \la{KTtrace}
\text{Tr}\, e^{-i \tilde{K}t}=\sum_{n=0}^\infty e^{-(n+1/2)t/2}=\frac{e^{-t/4}}{1-e^{-t/2}}.
\end{equation}
As expected from \nref{ModBde} and the cyclicity property of the trace this is independent of $\alpha$. However, we need a non-zero $\alpha$ to have a well defined computation. 
%
%
In fact, if we set $\alpha =0$ from the beginning and we compute the trace using the eigenstates \nref{Eigfphi} we get 
\begin{equation}
\la{wavetrace}	
\text{Tr}\, e^{-i K t}=\int d\omega \bigg(\bra{\omega_{+}}e^{-i K t}\ket{\omega_{+}}+\bra{\omega_{-}}e^{-i K t}\ket{\omega_{-}}\bigg)=4\pi \delta(0)\delta(t)
\end{equation}
where we used that $\bra{\omega'_{\pm}}\ket{\omega_{\pm}}=\delta(\omega'-\omega)$ and integrated over $\omega$ to obtain the delta function in time. This divergent result is due to the continuous spectrum of $K$. 
In fact, it is easier to understand in the $\phi$ coordinates discussed around \nref{NorPhi}. In those coordinates the eigenvectors are plane waves and the trace can be defined by putting a large volume cutoff $V_\phi$ in the $\phi$ coordinates. 
Then $\delta(0)$ factor in (\ref{wavetrace}) would be interpreted as $V_{\phi}/(2 \pi)$ and 
%
\begin{equation}
\text{Tr}\,e^{-i K t}=2 V_{\phi} \delta(t).
\end{equation}
However, we can get other answers for $\text{Tr}\,e^{-i K t}$ if we choose a different basis for the trace. For example, let us consider the overcomplete basis involving coherent states  
\begin{equation}
\ket{\lambda}=e^{-|\lambda|^{2}/2}e^{\lambda a^{\dagger}}\ket{0} ~,~~~~{\rm with } ~~~~\mathbb{I}=\frac{1}{\pi}\int d^{2}\lambda \ket{\lambda} \bra{\lambda}.
\end{equation}
The matrix elements are, see \nref{MelK} in appendix \ref{CohApp},
\begin{equation}
\bra{\lambda}e^{-i K t}\ket{\lambda}=\frac{1}{\sqrt{\cosh t/2}}e^{\frac{(\lambda e^{-t/4}+\bar{\lambda}e^{t/4})^{2}}{2\cosh t/2}-\frac{1}{2}(\lambda+\bar{\lambda})^{2}},
\end{equation}
and we find 
\begin{equation}
\la{coherenttrace}
\text{Tr}\, e^{-i K t}=\frac{1}{\pi}\int d^{2}\lambda \bra{\lambda}e^{-i K t}\ket{\lambda}=
\frac{e^{-t/4}}{1-e^{-t/2}}
\end{equation}
where we did the  gaussian integral of $\lambda$, which is well defined for all $t > 0$. This agrees with \nref{KTtrace}.  
However, doing the integral that led to this in different orders would give different answers. 
For example, in obtaining \nref{coherenttrace} we first did an intermediate $x$ integral \nref{diaglamb} to compute the overlap. If one instead did the $\lambda$ integral from the trace first, one would obtain a term in the $x$ integral proportional to
\begin{equation}
\delta\bigg(x \sinh \frac{t}{4} \bigg)=\frac{1}{|x|} \delta(t)+\frac{1}{|\sinh \frac{t}{4}|}\delta(x)
\end{equation}  
and each of these terms would turn out to give one of the different answers we obtained. To be precise, the first one would give rise to $\nref{wavetrace}$, and the second one would give a rise to $\nref{coherenttrace}$, leading to a total answer that is the sum of them.

The conclusion from this is that $\textrm{Tr}[e^{ -i K t} ]$ is ill defined and its answer depends on the regularization method that is used to define it. Only in some of them we reproduce \nref{KTtrace}. 

 Similar traces and their regularization have also been studied in \cite{Anninos:2020hfj,Sun:2020ame} in a different context.

\subsection{Exposing the non-unitarity of the $\tilde{K}$ evolution}


Since $K$ is a Hermitian operator, the evolution operator $e^{-i K t}$ is unitary and preserves the norm. However, since $\tilde{K}$ is non-Hermitian one expects $e^{-i\tilde{K}t}$ to   decrease the norm since all its eigenvalues have  negative imaginary part. As all the eigenvalues of $\tilde K$ are non-zero, one might naively expect that the norm decreases relatively fast. However, we will see that the situation is   more subtle. 

We can calculate the norm by expanding a general quantum state $\ket{\psi}$ in terms of right eigenstates
%
%
\begin{equation}\label{normKtilde}
\left|e^{-i \tilde{K}t}\ket{\psi}\right|^{2}=\sum_{n,m}c_{m}^{*}c_{n}\bra{m_{r}}\ket{n_{r}} \times e^{-(n+m+1)t/2}.
\end{equation}
From this expression, one might incorrectly conclude that the norm should decrease faster than $e^{ -t/2}$. This conclusion would be correct if the right eigenstates were all orthogonal and each term in (\ref{normKtilde}) comes with positive coefficient. However, they are not, so this is a sum of terms with complex coefficients with different phases, 
and therefore this sum does not necessarily decrease.\footnote{ 
A simple toy example is the fuction $   2/( 1 + e^{-t} )$ which can be expanded in powers of $e^{-t}$, but is not a decreasing function. } Another reason for this to fail is that the sum might not be convergent for small $t$.  
 
As an explicit example, let us compute  (see \nref{NotKte} in appendix \ref{CohApp}) 
\begin{equation} \la{NorKz}
\left|e^{-i \tilde{K} t}\ket{0}\right|^{2}=\frac{1}{\sqrt{1+\sinh t \sin \alpha+(\cosh t-1)\sin^{2}\alpha}}
\end{equation}
For small $\alpha$,  this norm remains one up to a time of order $t \sim \log { 1 \over \alpha}$, where it starts decreasing. 
In fact,  for small $\alpha $ we can approximate \nref{NorKz} as 
\begin{equation}
\label{alpnorm}	
	\left|e^{-i \tilde{K} t}\ket{0} \right|^{2} \approx e^{-t/2}\frac{1}{\sqrt{\frac{\alpha}{2}+e^{- t}}}=\sqrt{\frac{2}{\alpha}}e^{-t/2}\sqrt{\frac{1}{1+\frac{2}{\alpha}e^{-  t}}}
\end{equation}
which implies that the expansion in terms of $\tilde K$ eigenstates only converges when $ { 2 \over \alpha } e^{ -   t } < 1$. 

This is an analogue of the general feature of black holes that the quasinormal mode expansions only converge for sufficiently large times \cite{Warnick:2022hnc,Ching:1995rt}, see appendix \ref{app:incomplete} for another example.

  \section{Conclusions }

In this paper,  we have mainly focused on understanding aspects of the double cone wormhole geometry introduced in \cite{Saad:2018bqo}.  
We have presented the full computation as a trace over the bulk Hilbert space of the double cone. This has the advantage of giving the precise normalization. For this purpose it is actually necessary to make an identification on one of the phase space variables that describes the empty wormhole, namely  $T_{\rm rel} \sim T_{\rm rel} + T$. So in this respect, it is not the same as the bulk hilbert space of the decompactified and familiar Lorentzian wormhole, where $T_{\rm rel}$ displays no identification. 

For the matter fields we discussed the 
  $i \epsilon$ prescription introduced in  \cite{Saad:2018bqo}.   
The wormhole has a boost symmetry generated by $K$. The   $i\epsilon $ prescription defines a new ``modified boost operator'' $\tilde K$, which generates a complexified geometry through its evolution. We explored a few properties of this modified boost. First,  we recalled that the eigenvalues of the modified boost operator are the quasinormal mode frequencies \cite{Bah:2022uyz}. This is valid for general black holes in general dimensions.  We discussed some applications of the modified boost. One was  the hydrodynamic contribution to the spectral form factor, reproducing \cite{Winer:2020gdp,Winer:2022gqz}. The other was the exponential suppression to the spectral form factor when we slightly change the couplings, reproducing the discussion in \cite{Cotler:2022rud}. 

We recalled that in $AdS_2$ we can view the modified boost as resulting from a conjugation by one of the generators, the generator $P$ that acts like spatial translations in the near horizon region. 
Inspired by that discussion, we proposed an algebraic construction for the general black hole in general dimensions which also involves a conjugation by an operator that acts likes translations in the near horizon region. In addition, we can continue to view $\tilde K = K - i \epsilon \hat H$, with $\hat H$ a positive operator that acts like time translations in the near horizon region. Using the proposed boundary construction in \cite{Leutheusser:2021frk} for the operators $\hat P$ and $\hat H$ we conclude that they should be viewed as two sided operators, defined in the large $N$ limit. 
This definition suggests that $\tilde K$ is a two sided operator coupling both sides.

 However, it is also possible to view the $i\epsilon $ prescription as resulting from a deformation that does not couple the two sides. In particular,    adding  fixed imaginary couplings we found in section \ref{ImCou} that backreaction leads to a deformed geometry with the right sign of the $\epsilon$ deformation.   In fact, this deformation  connects the double cone wormhole with the usual Euclidean wormholes, such as \cite{Giddings:1987cg} which involve imaginary 
 fields.\footnote{We are grateful to Douglas Stanford for valuable comments on these points.}

 Notice that the Saad-Shenker-Stanford $i \epsilon $ prescription can be defined in any situation with a horizon, including the static patches of de-Sitter space. In that case, it leads to the de-Sitter quasinormal modes. In particular, this prescription tells us how to view the quasinormal modes as the eigenvalues of a non-hermitian second quantized operator. The operator $\tilde K$ that corresponds to time evolution in the complexified geometry.  
 
   For the special case of JT gravity plus matter we have considered the effects of backreaction due to matter fields. This is an effect that becomes very small for large times $T\gg \beta(E)$. Nevertheless, we have found that the matter deforms the geometry in a complex direction which is the {\it opposite} from the one produced by  the $i \epsilon$ regularization needed to make the matter action well defined. A similar feature was observed in \cite{Cotler:2022rud} in the case of the ramp with different couplings on the two sides. 
      
   We have also discussed a toy model for $K$ and $\tilde K$ that is even simpler than what happens in $AdS_2$ and is helpful for understanding how the ``small'' $i \epsilon$ modification of $K$ can lead to dramatic changes in the eigenvalues, but small changes in the the matrix elements of simple states.

\subsection*{Acknowledgments}

We particularly thank D. Stanford for insightful comments and for correcting some of our  misconceptions. We also would like to thank D. Jafferis, S. Leutheusser, H. Lin, H. Liu, P. Saad,  S. Shenker, Z. Sun, G. Turiaci,  E. Witten and Z. Yang for discussions.   

J.M. is supported in part by U.S. Department of Energy grant DE-SC0009988.   

\appendix

\section{The Kontsevich-Segal  criterion and the decay of correlators}\label{subsec:kscrit}

The connection between the decay of correlators and the  complexified metrics can be made more precise by using the Kontsevich-Segal criterion    \cite{Kontsevich:2021dmb,Witten:2021nzp}. This criterion states that the complexified
 metric should be such that 
\begin{equation} \la{KSc}
\text{Re}[\sqrt{g}g^{i_{1} j_{1}}...g^{i_{p}j_{p}}F_{i_{1}i_{2}...i_{p}}F_{j_{1}j_{2}...j_{p}}]>0
\end{equation}
where $F$ is any $p$ form. This implies that we get a convergent path integral similar to the ones we get in Euclidean space. Therefore, in a spacetime with a time translation symmetry,
one would expect it to lead to decay of correlators at long times. The complexified metric \nref{BHme} or \nref{NHmec} does indeed obey this criterion \cite{Witten:2021nzp}. Let us first review this argument. 

We define a complexified Lorentzian solution to be forward moving in time if 
\begin{equation}
\sqrt{g}=i\sqrt{-g} 
\end{equation}
with the motivation that this property would take one from the Euclidean path integral to the usual Lorentzian path integral with positive time evolution.\footnote{Solutions with $\sqrt{g}=-i \sqrt{-g}$ would be analogously backwards moving and have similar properties with appropriate minus signs} This implies the KS criteria \nref{KSc} for forward moving backgrounds can also be stated as 
\begin{equation}
\label{fwrdmv}
	\text{Im}[\sqrt{-g}g^{i_{1} j_{1}}...g^{i_{p}j_{p}}F_{i_{1}i_{2}...i_{p}}F_{j_{1}j_{2}...j_{p}}]<0
\end{equation} 
Then  \nref{fwrdmv} implies 
\begin{equation}
	 \text{Im}[\sqrt{-g}]=\text{Im}[\sinh(\tilde \rho-i \epsilon)]=-\cosh \tilde \rho \sin \epsilon  <0 ~,~~~~~{\rm or }~~~~~ \epsilon>0
\end{equation}

We want to show that time evolution in backgrounds respecting \nref{fwrdmv} is decaying, e.g, the eigenvalues of its time evolution generator have negative imaginary part. To see this explicitly, let us study a generic generator of time translations $\tilde{K}$ for scalar field in this background, with Lorentzian   action
\be 
S = \int  d^D x \sqrt{-g} \left( - \half  g^{\mu \nu } \partial_\mu \phi \partial_\nu \phi   - V(\phi) \right) 
\ee 
In terms of the momentum  defined as 
\be 
\pi =   \sqrt{-g} (- g^{tt} \partial_t\phi )
\ee
we can write the generator of $t$ translations as:
\be
	\tilde K =\int d^{D-1}x (\pi \partial_{t}\phi-\sqrt{-g}\mathcal{L})=\frac{1}{2}\int d^{D-1}x\bigg(\pi^{2}(-\sqrt{-g}g^{t t})^{-1}+\sqrt{-g}g^{i j}\partial_{i}\phi \partial_{j}\phi+\sqrt{-g}V(\phi)\bigg)
\ee

From here, proving the imaginary part of $\tilde K$ is negative is straightforward. Namely, the Konsevich-Segal condition implies that both $\sqrt{-g}g^{t t}$, and $\sqrt{-g}g^{i j}\partial_{i}\phi \partial_j \phi$ 
have negative imaginary part 
, so if $V(\phi)>0$, $\text{Im}\, \tilde K <0$ holds as an operator statement as long as $\phi$ and $\pi$ are hermitian operators.\footnote{One might worry about negative potentials. We will give a better argument for the positivity of $\tilde K$ in section \ref{sec:algebra}.}

\section{Incompleteness of quasinormal modes}\label{app:incomplete}

\begin{figure}[h]
	\begin{center}
		\includegraphics[scale=.25]{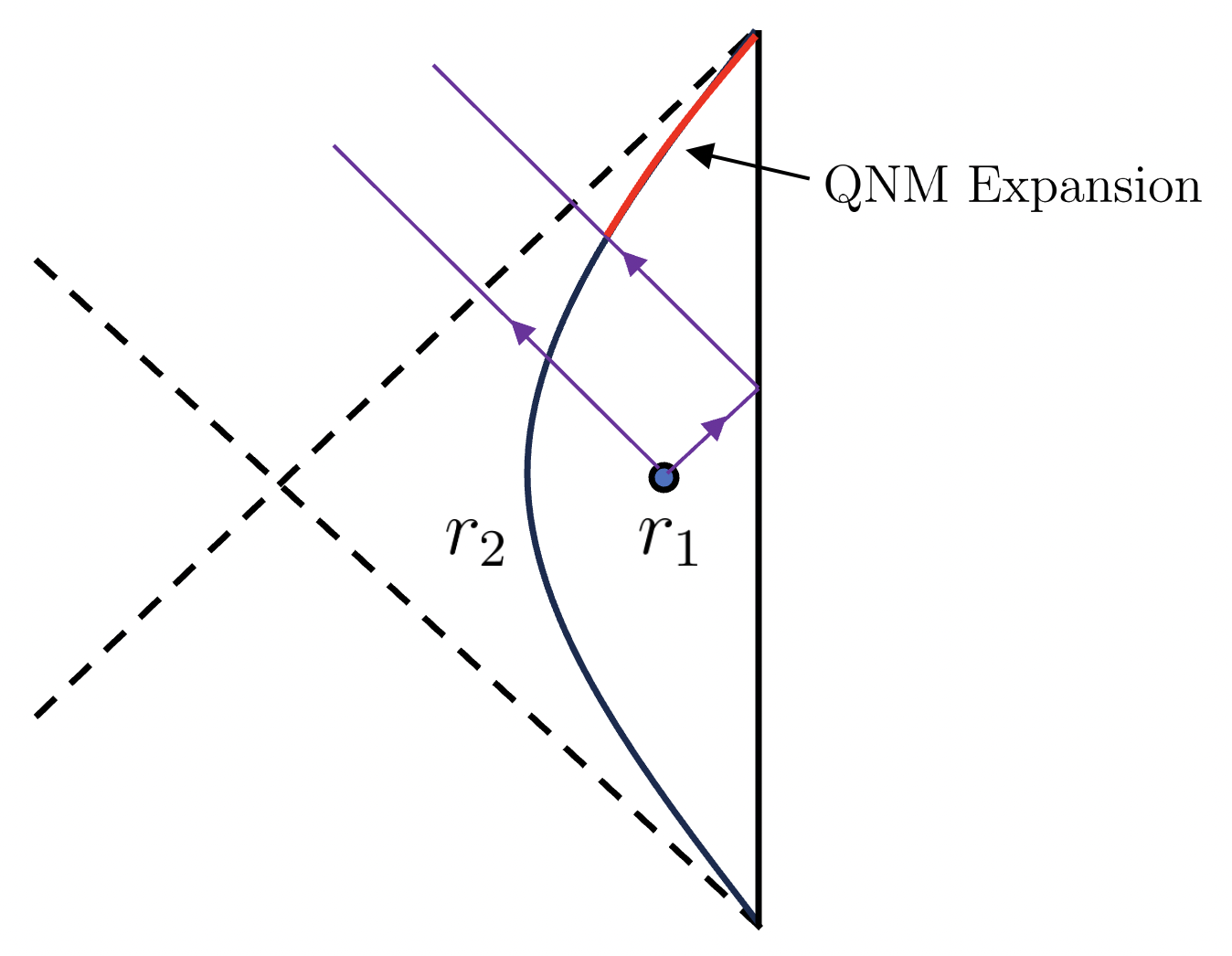}
	\end{center}
	\caption{We see that the QNM expansion is only valid at suitably late times, times in the red region.   }
	\label{QNMExpansion}
\end{figure}

One could wonder based on results such as \nref{compads2} and \nref{CompM} if we can decompose an arbitrary solution in terms of quasinormal modes. The answer is that it is not always possible, but it is possible at late times. To give an example, we can go back to the massless scalar field in AdS$_2$ and consider the following correlator
%
%
%
\begin{equation}
	\langle \phi(r_2,t) \partial_t \phi(r_1,0) \rangle \propto \bigg(\frac{r_{1}r_{2}}{e^{t}-r_{1}r_{2}}-\frac{r_{1}r_{2}}{e^{-t}-r_{1}r_{2}}+\frac{r_{1}}{r_{1}-e^{t}r_{2}}-\frac{r_{1}}{r_{1}-e^{-t}r_{2}}\bigg) 
\end{equation}
%
where we are using the coordinate system in sec. \ref{QNMExpl}. We can expand this in powers of $e^{-t}$ only when 
\be 
e^{ t} |r_1 r_2| > 1 ~,~~~~~~ e^t {|r_2| \over |r_1| } > 1 ~,~~~~~e^t { |r_1| \over |r_2| } > 1 
\ee 
This happens when the point $(r_2 ,t)$ is to the future of the light rays that come from $r_1$, both directly and reflecting from the boundary, see figure \ref{QNMExpansion}. This means that this correlator can be expanded in quasinormal modes only for times $t$ which are sufficiently large. This should be contrasted with the eigenbasis of $H$, where we would be able to expand a similar correlator for any positive time. 

If one wants the correlator above to make sense for any point in the slice, the condition $|r|_{min}^{2}e^{t}>1$ is sufficient since $|r|<1$ for any point not on the boundary. $r_{min}$ is the smallest distance of points in the slice to $r=0$, which is of order $\alpha$ for small $\alpha$, so the correlator can be expanded in quasinormal modes in the entire slice for $t >\frac{2}{\alpha}\ln \alpha^{-1}$.



\section{Explicit average over couplings in JT gravity plus matter} 

\la{AveAppendix}

In this appendix we consider JT gravity plus matter. We want to introduce an explicit average over couplings by averaging over the boundary conditions of the matter fields. In other words, we average over explicit couplings that we have in the gravity description. 
Our goal is to see whether such an average leads to a correction for the gravitational saddle point for a wormhole that would be equivalent to the   $ - i \epsilon$ prescription of \cite{Saad:2018bqo}. We also assume that we have a large number of matter fields to make a controlled computation. Each matter field has its own independent boundary condition that we average over.  We will suppress the index on the matter fields, they are assumed throughout the discussion below. 

\subsection{Average over physical couplings } 
\la{AveReal} 

We consider a field propagating in $AdS_2$ and we consider the effects of averaging over its boundary conditions. This is equivalent to inserting operators at the boundary. 
In other words, by averaging the coupling $g$ with a gaussian measure, we have the following correction to the wormhole 
 \bea 
 &~&~ \int d g e^{ - \half {g^2 \over \sigma^2 } } \langle \exp\left( i g \int du_L O_L(u_L) - i g \int du_R O_R(u_R)  \right) \rangle = \la{TwoSC}
 \\ 
  &~&~ = 
   \left\langle \exp\left(   - \half { \sigma^2 } \left[ \int du_L O_L(u_L) -   \int du_R O_R(u_R) \right]^2 \right) \right\rangle
  \cr 
  &~&~
  \approx \exp\left( - \half \sigma^2 \left[  \int du_L du'_L \langle \bar T O_L O_L \rangle +  \int du_R du'_R \langle T O_R O_R \rangle -  2  \int du_L du'_R \langle O_L O_R \rangle  \right]   \right) ~~~~~~~~~ \la{expexpv}
 \eea 
  where $T$ is time ordering and $\bar T$ is anti-time ordering.  We have also assumed that the $O$ are dual to generalized free fields. 
 We can use 
 \be \la{COrrRR}
 \langle T O_R(u) O_R(u') \rangle  \propto { 1 \over \left[ \beta i \sinh ({ \pi   |u_{12}| - i\epsilon \over \beta} )\right]^{2 \Delta }  } ~,~~~~~~~~~~~~
 \langle O_L(u) O_R(u')\rangle \propto { 1 \over \left[ \beta \cosh ({ \pi   u_{12} \over \beta} )\right]^{2 \Delta }  } 
 \ee 
 It turns out that this then leads to a cancellation between all the terms in  \nref{expexpv}.  The cancellation can be seen by combining the RR and LL contributions into a single integral over the argument of the $\sinh$ in \nref{COrrRR} and by a contour deformation argument we can relate this integral to one from the RL and LR contributions. 
 So that this average over couplings does not lead to any effect. 
 
 We can reach the same conclusion more quickly in some special cases. For example, consider a bulk massless bosonic field with Dirichlet boundary conditions. In this case the change in boundary conditions only affects the constant value of the field in the bulk, which does not lead to any further effect. 
 
 In fact, the average over couplings treated in this approximation is the same as adding fixed couplings since if we had fixed couplings we would add 
 \bea \la{CoupCo}
 &~ & \left\langle \exp\left(  - i \hat g \int du_L O_L(u_L)  + i \hat g \int du_R O_R(u_R) \right)  \right\rangle 
 \cr 
 &~& 
 = \exp\left( - \half {\hat g}^2 \left[  \int du_L du'_L \langle \bar T O_L O_L \rangle +  \int du_R du'_R \langle T O_R O_R \rangle -  2  \int du_L du'_R \langle O_L O_R \rangle  \right]\right)~~~~
 \eea 
 which has the same form as \nref{expexpv}. This explains partly why by averaging over coupling we are not getting any effect.

 \subsection{Average over couplings including backreaction } 
 
 In the above discussion we have neglected the possible backreaction of the matter fields on the geometry. Taking backreaction into account, we actually find a non-vanishing effect, but it deforms the geometry in the opposite direction, i.e. the direction that is oppsite to the desirable direction for modified boost. 
 
  We use the following ansatz for the  the Schwarzian variables describing  $t_R(u)$ and $t_L(u)$ with 
  \bea 
  t_R(u) &= & b { u \over T - i \beta } ~,~~~~~~~ u \sim u + T - i \beta 
  \cr 
  t_L(u) &=&  b { u \over T + i \beta } ~,~~~~~~~ u \sim u + T + i \beta 
  \eea
  where $t \sim t + b$.  Here $t$ can be viewed as the bulk time in the standard coordinates \nref{AdS2Me}.   
 The correlators can be computed from 
 \be 
  \langle O(u) O(u') \rangle = \left( { t'(u) t'(u') \over [ i\sinh \frac{t(u) - t(u')}{2}]^2 } \right)^{\Delta } ,
  \ee 
 which give
  \bea 
   \langle T O_R(u) O_R(u') \rangle &=& b^{2\Delta } { 1 \over (T- i \beta)^{2\Delta } }  {1  \over [i \sinh \frac{\tau}{2}     ]^{2\Delta }   } ~,~~~~~\tau = { b |u-u'| \over T - i \beta } ,
   \cr 
  \langle \bar T O_L(u) O_L(u') \rangle &=& b^{2\Delta } { 1 \over (T+ i \beta)^{2\Delta } }  {1  \over [ (-i)\sinh \frac{\tau}{2}      ]^{2\Delta }   } ~,~~~~~\tau = { b |u-u'| \over T + i \beta }  ,
   \cr 
  \langle \bar T O_L(u) O_R(u') \rangle &=& b^{2\Delta } { 1 \over (T^2+  \beta^2)^{\Delta } }  {1  \over [ \cosh \frac{\tau}{2}      ]^{2\Delta }   } ~,~~~~~\tau =  b \left(  {u  \over T - i \beta } - { u' \over T+ i \beta } \right).~~~~
  \eea 
   We now use 
   \bea \la{IntVa}
I_{RR} &=&    2 \int_0^\infty  { 1 \over [ i  \sinh (\frac{|t|}{2} - i \epsilon)  ]^{2 \Delta } }= { 2 e^{ - i \pi \Delta }  \over \cos \pi \Delta } \, { \Gamma(\Delta) \sqrt{\pi } \over  \Gamma(\half + \Delta)     } ,
\cr 
I_{LL} &=&   2 \int_0^\infty  { 1 \over [ (-i ) \sinh (\frac{|t|}{2} + i \epsilon)  ]^{2 \Delta } }= { 2e^{  i \pi \Delta }  \over \cos \pi \Delta } \, { \Gamma(\Delta) \sqrt{\pi } \over  \Gamma(\half + \Delta)     } ,
 \cr 
2 I_{LR} &=&     2 \int_{-\infty}^\infty { dt \over [ \cosh \frac{t}{2}]^{2\Delta } } = { 4 \sqrt{\pi} \Gamma(\Delta) \over \Gamma( \half + \Delta ) } ,
\eea 
 where we assumed that $ 0  < \Delta < \half $ so that we do not have to worry about what happens near $t\sim 0$. 
   
   Now the contribution to the partition function from the average over couplings takes the form, from \nref{expexpv}, 
   \bea 
   \log Z_{\rm ave} &=& - N_f \sigma^2  b^{ 2 \Delta -1 } {\Gamma(\Delta)  \sqrt{\pi } \over \Gamma(\half + \Delta )  }  \left[ { 
   (T- i \beta)^{ - 2 \Delta + 2} e^{ - i \pi \Delta }   + (T+ i \beta)^{ - 2 \Delta + 2} e^{  i \pi \Delta }  \over \cos \pi \Delta } - 2 (T^2 + \beta^2)^{-\Delta+1 } \right] ~~~~~~~~~~~
   \eea
   where $N_f$ is the number of matter fields in the bulk. 
   Notice that one of the $u$ integrals gives us a factor of $T\pm i \beta$. The second is converted to an integral over $t$ at the expense of a factor of $(T \pm i \beta) \over b $. This is the origin of the factors in the above equation. 
   
   We can expand the above expression to first order in $\beta $ to obtain 
   \be \la{ZaveAn}
   \log Z_{\rm ave}  \sim  N_f  \sigma^2  b^{ 2 \Delta -1 } T^{2 - 2 \Delta } { \beta \over T } {\Gamma(\Delta)  \sqrt{\pi } \over \Gamma(\half + \Delta )  }  ( 1- \Delta) { \sin \pi \Delta \over \cos \pi \Delta } 
    \ee 
    To this, we should add some terms that come from the Schwarzian and from the original matter partition function. These are 
    \be 
    \log Z_{\rm rest} = -  { \phi_r b^2 \beta \over T^2 } + \log Z_m (b)
    \ee 
  Extremizing this with respect to $b$ gives   
    \be \la{AEst}
   0= \partial_b [ \log Z_{\rm rest} + \log Z_{\rm ave} ] = - 2 { \phi_r b \beta \over T^2 } - {\cal E}(b) -   C b^{ 2 \Delta - 2 } T^{2 - 2 \Delta } { \beta \over T }  
    \ee 
    where 
    \be 
    C = 2 \sigma^2 \left(\half - \Delta \right)(1-\Delta) { \sin \pi \Delta \over \cos \pi \Delta } {\Gamma(\Delta)  \sqrt{\pi } \over \Gamma(\half + \Delta )  }
    \ee 
    we see that for $0 < \Delta < \half $, $C>0$, so that the solution has $\beta < 0$ as it was in \nref{saddeqn}, when $C=0$. 
     this pushes $\beta$ in the wrong direction. In the range $0 < \Delta < 3/2 $ this correction is more important, for large $T$ than the Schwarzian term (the first in \nref{AEst}).

 \subsection{Complex couplings } 
 
 \la{AveComp}
 
 We get a different result if we allow the couplings to be complex and we take the coupling on the left side to be the complex conjugate of the one on the right. Then,   
  instead of \nref{TwoSC}, we get 
 \bea 
 &~&~ \int d^2 g\, e^{ -   { g \bar g  \over \sigma^2 } } \langle \exp\left( i g \int du_L O_L(u_L) - i \bar g \int du_R O_R(u_R)  \right) \rangle \la{TwoSCcc}
 \\ 
  &~&~ = \exp\left(  \sigma^2    \int du_L du'_R \langle O_L O_R \rangle    \right) ~~~~~~~~~ \la{expexpcc}
 \eea 
 We see that the same side cases drop out. This leads to an ``attractive'' force between the two sides of the wormhole. The simplest way to see it is to note that the real part of the couplings is the same as in section \ref{AveReal}. The imaginary parts are opposite on the two sides and imaginary. Then we get an effect similar to the one discussed in \cite{Cotler:2022rud}, but with the opposite sign. In \cite{Cotler:2022rud}, as in our discussion in section \ref{sec:newsad} the wormhole is shifted by $i \epsilon $ in the ``wrong'' sign. With the averages over complex couplings we are discussing here it is shifted in the ``right'' direction. In fact,  \nref{expexpcc}   produces a negative contribution to the energy 
  ${\cal E}$  of section \ref{sec:newsad}. If we average over enough couplings so that the total sign of the bulk energy ${\cal E}$ gets flipped, then we will have a well defined wormhole following the discussion of sections \ref{sec:newsad}, \ref{sec:local} since now ${\cal E} < 0$ and then $\beta $ will be positive in equation \nref{saddeqn}.

  We can also consider fixed complex couplings.  A computation similar to \nref{CoupCo} tells that we need to consider 
  \bea \la{intex2ptA}
   \exp\left(  \half h^2 \left[  \int du_L du'_L \langle \bar T O_L O_L \rangle +  \int du_R du'_R \langle T O_R O_R \rangle +  2  \int du_L du'_R \langle O_L O_R \rangle  \right]\right)~~~~
 \eea 
 where the coupling is $\hat g = i h$ with real $h$. Using \nref{IntVa} we get 
 \be 
 \log Z_{\rm couplings} \sim N_f   h^2 b^{2 \Delta -1} T^{2 -2\Delta } { \Gamma(\Delta ) \sqrt{\pi } \over \Gamma(\half + \Delta ) } 
 \ee 
 
 For $ 0 < \Delta < \half $ this does not help with the sign of the $i \epsilon$ deformation. But for $\Delta >\half$, the derivative with respect to $\Delta $ changes sign and it can help in producing the desired sign of the $i \epsilon $ deformation. 
 In fact, for $\Delta =1$ we could get a purely Euclidean wormhole just from this term and the the JT gravity contribution \cite{Garcia-Garcia:2020ttf}. 
 It is curious that there is a different behavior depending on the value of $\Delta$. Notice that $ \Delta < 1/2$ is the range where the perturbation is such that there is no UV divergence when we consider the integral of the two point function of the operator as in \nref{intex2ptA}.

 \section{Analogue of the modified boost operator for open systems}
 
The quasinormal mode decay is a generic feature of the dissipative dynamics in open quantum systems. In our case, the quantum system can be viewed as the quantum fields outside the black hole, outside a small distance from the horizon, $\rho \geq \epsilon$. It interacts with the quantum fields very close to the horizon, $|\rho | < \epsilon$. So we can view this as a special case of a dissipative open quantum system. In gravity, the modified boost operator can be viewed as describing the effective dynamics of this system and it involves the second copy, the other side of the black hole. When we consider the modified boost we avoid the $ |\rho |< \epsilon $ direction by introducing a complex path between the left side and the right side. We can view this as a complex coupling between two copies of the quantum system describing the fields for $|\rho| > \epsilon$.  
 One can wonder whether a similar construction of the ``modified boost operator" exists for general open quantum systems.\footnote{We thank Leonard Susskind for discussion on this point.} 

In fact something somewhat similar arises for general systems as follows.  We can reinterpret the two sides of a Schwinger Keldysh contour for a density matrix as two quantum systems, and we would generate couplings between them when we trace out the environment. More explicitly, this can be seen as follows (see for example \cite{Prosen:2012sn,Zhou:2023qmk}). The dynamics of an open quantum system can be modeled by the Lindblad equation
\begin{equation}\label{Lind}
	\frac{d\rho}{dt} = - i[H,\rho] + \sum_{i} \gamma_i\left(2 L_i \rho L_i^\dagger - \{L_i^\dagger L_i,\rho\} \right),
\end{equation}
where $L_i$ are the Lindblad jump operators and $\gamma_i$ denote the dissipation strength. We can map the density matrix (and also other operators) to a state in a two sided Hilbert space via the Choi–Jamiołkowski isomorphism, i.e. $\rho (t)\rightarrow \ket{\psi(t)} = \sum_{ij} \rho_{ij}(t) \ket{j}_L\ket{i}_R$, where $\{\ket{i}\}$ is an orthonormal basis of states. The Lindblad equation (\ref{Lind}) is then equivalent to a Schrodinger evolution in the two sided Hilbert space with a Hamiltonian
\begin{equation}
	H_{\textrm{Lindblad}} = H_R - H_L^T -i \sum_{i} \gamma_i \left[ -2 L_{i,L}^* L_{i,R}  + L_{i,R}^\dagger L_{i,R} + (L_{i,L}^\dagger L_{i,L})^* \right].
\end{equation}
The Hamiltonian $H_{\textrm{Lindblad}}$ is non-Hermitian and couples the two sides. Note that this discussion is similar to the black hole case but not exactly the same, since here the density matrix $\rho(t)$ is \emph{not} given by the reduced density matrix of $\ket{\psi(t)}$ by tracing out one side.

\section{Gauge invariant generators}\label{app:gaugeinv}

In this appendix we review the formalism of gauge invariant operators for JT gravity introduced in \cite{Lin:2019qwu}. The idea is that by dressing the matter $SL(2)$ generators $Q_{m}^{a}$ appropriately
\begin{equation}
\label{gaugeinvgen}
	G^{A}=e_{a}^{A}Q_{m}^{a}
\end{equation}
with $e_{a}^{A}=e_{a}^{A}(X_{r},X_{l})$ functions of the boundary coordinates, the resulting generator will not change under a simultaneous gauge transformation of both matter and boundaries degrees of freedom, which is a gauge redundancy. A particular set of such $e_{a}^{A}$ can be found explicitly to be given by
\begin{equation}
	e_{a}^{0}=\frac{\epsilon_{a b c}X_{l}^{b}X_{r}^{c}}{X_{l}\cdot X_{r}} \text{, } e_{a}^{-1}=-\frac{X_{r a}+X_{l a}}{\sqrt{-2 X_{l}\cdot X_{r}}} \text{, }e_{a}^{1}=\frac{X_{r a}-X_{l a}}{\sqrt{-2 X_{l}\cdot X_{r}}}
\end{equation}

To understand how the generator of time translations $\partial_{t}$ is expressed in terms of \nref{gaugeinvgen} is convenient to work in embedding coordinates, where positions in the modified $AdS_{2}$ spacetime are given by
\begin{equation}
	\begin{gathered}
		\label{Yeq}
		Y=(Y^{-1},Y^{0},Y^{1})=(\cosh \rho, \sinh \rho \sinh t, \sinh \rho \cosh t)\\
	\end{gathered}    
\end{equation}
with $\rho=\tilde{\rho}-i \alpha$, and $\tilde{\rho}$ and $\alpha$ are real. One defines $X_{r,l}$ by re-scaling the above by $e^{\tilde{\rho}_{c}}/2$, where $\tilde{\rho}_{c}$ is a big positive real number standing for real part of the cutoff radius of the right boundary. In terms of that $\rho_{r}=\tilde{\rho}_{c}-i \alpha$ and $\rho_{l}=-\tilde{\rho}_{c}-i \alpha$ with the coordinate format the same as in \nref{Yeq}, leading to
\begin{equation}
	\begin{gathered}
		X_{r}^{a}=e^{-i \alpha}(1,\sinh t,\cosh t)\\
		X_{l}^{a}=e^{i \alpha}(1,-\sinh t,-\cosh t)
	\end{gathered}    
\end{equation}
and putting this all together one can find the $e_{a}^{A}$ in this background to be
\begin{equation}
	\begin{gathered}
		e_{a}^{0}=(0,\cosh t,-\sinh t) \text{, } e_{a}^{-1}=(\cos \alpha,-i \sin \alpha \sinh t,i \sin \alpha \cosh t)\\
		e_{a}^{1}=(i \sin \alpha,-\cos \alpha \sinh t, \cos \alpha \cosh t)
	\end{gathered}    
\end{equation}
Setting $t=0$, in the spirit of taking $t_{l}=t_{r}=0$ as in \cite{Lin:2019qwu}, one would obtain
\begin{equation}
\begin{gathered}
	G^{-1}=\cos \alpha \, Q_{m}^{-1}+i \sin \alpha \, Q_{m}^{1}\\
	G^{1}=\cos \alpha\, Q_{m}^{1}+i \sin \alpha\, Q_{m}^{-1} \text{, } G^{0}=Q_{m}^{0}
\end{gathered}	
\end{equation}
The crucial step here is to recognize that $Q_{m}^{-1}$ is the generator of time translations in this background, which is $\tilde{K}$. Therefore, one can rewrite $\tilde{K}$ in terms of gauge-invariant operators
\begin{equation}
	\tilde{K}=Q_{m}^{-1}=\cos \alpha\, G^{-1}-i \sin \alpha\, G^{1}=\cos \alpha\, K-i \sin \alpha\, H
\end{equation}
where in the last line we wrote $G^{-1}=K$ and $G^{1}=H$ since they are the gauge invariant generalization of these generators (we denoted them by $G_K, G_H$ in (\ref{MOdBo})).

\section{Some details on the toy model}
\la{ToyApp}

The explicit forms of the left and right eigenstates are 
\begin{equation}
\begin{gathered}
	\la{lrwvfct}
     \psi_{n,r}(x)=\frac{e^{\frac{1}{2}\cot \alpha\, x^{2}}}{({2\cos^{2}\frac{\alpha}{2}})^{\frac{1}{4}}}\bigg(\tan \frac{\alpha}{2}\bigg)^{\frac{n}{2}}\psi_{n}\bigg(\frac{x}{\sqrt{\sin \alpha}}\bigg) \text{, } ~~~~~~  \psi_{n,l}(x)=\frac{e^{-\frac{1}{2}\cot \alpha\, x^{2}}}{({2 \sin^{2}\frac{\alpha}{2}})^{\frac{1}{4}}}\bigg(\cot \frac{\alpha}{2}\bigg)^{\frac{n}{2}}\psi_{n}\bigg(\frac{x}{\sqrt{\sin \alpha}}\bigg)
\end{gathered}     
\end{equation}
with $\psi_{n}$ the wavefunction of the usual harmonic oscillator, that is
\begin{equation}
\psi_{n}(x)=\sqrt{\frac{1}{2^{n}n!\sqrt{\pi}}}e^{-\frac{x^{2}}{2}}H_{n}(x)
\end{equation}
where $H_{n}(x)$ is the nth Hermite polynomial. The completeness relation therefore follows immediately since
\begin{equation}
\begin{gathered} \la{Compleness}
    \sum_{n}\psi_{n,r}(x)\psi_{n,l}(x')=\sum_{n}\frac{e^{\frac{1}{2}\cot \alpha (x^{2}-x'^{2})}}{\sqrt{\sin \alpha}} \psi_{n}\bigg(\frac{x}{\sqrt{\sin \alpha}}\bigg)\psi_{n}\bigg(\frac{x'}{\sqrt{\sin \alpha}}\bigg)=\delta(x-x')
\end{gathered}    
\end{equation}

We can compute the overlap $\bra{n_{r}}\ket{0}$ as:

\begin{equation}
\begin{gathered}
    \bra{n_{l}}\ket{0}=\int \frac{dx}{\sqrt{2^{n}n!\pi}(\sin \alpha)^{1/4}}\bigg(\cot \frac{\alpha}{2}\bigg)^{\frac{n}{2}+\frac{1}{4}}e^{-\frac{1}{2}x^{2}-\frac{1}{2}\cot\big(\frac{\alpha}{2}\big)x^{2}}H_{n}\bigg(\frac{x}{\sqrt{\sin \alpha}}\bigg)\\
    = \frac{(\sin \alpha)^{1/4}}{\sqrt{2^{n}n!\pi}}\bigg(\cot \frac{\alpha}{2}\bigg)^{\frac{n}{2}+\frac{1}{4}}\int dx e^{-\frac{(1+\cos \alpha+\sin \alpha)}{2}x^{2}}H_{n}\big(x\big)
\end{gathered}    
\end{equation}
and one can use the identity:

\begin{equation}
    \int dx e^{-\gamma x^{2}} H_{2m}(x)=\frac{(2m)!}{m!}\bigg(\frac{1}{\gamma}-1\bigg)^{m}\sqrt{\frac{\pi}{\gamma}}
\end{equation}
to obtain the answer, for $n$ even:

\begin{equation}
    \bra{2 m_{l}}\ket{0}=\frac{(-1)^{m}(\sin \alpha)^{1/4}\sqrt{2(2m)!}}{2^{m}m!\sqrt{1+\sin \alpha+\cos \alpha}}\bigg(\frac{\cos \alpha+\sin \alpha-1}{\cos \alpha+\sin \alpha+1}\bigg)^{m}\bigg(\cot \frac{\alpha}{2}\bigg)^{m+\frac{1}{4}}
\end{equation}
once one obtains this, it's straightforward to evaluate the overlap of the evolved state with the original one, to then obtain the return probability, that is:

\begin{equation}
    \bra{0}e^{-i \tilde{K}t}\ket{0}=\sum_{m}\bra{0}\ket{2m_{r}}\bra{2m_{l}}\ket{0}e^{-t/4}e^{-m t}
\end{equation}
and one can proceed quickly by noticing the left overlap is just the right overlap with $\alpha \rightarrow \pi-\alpha$, leading to:

\begin{equation}
\begin{gathered}
    \bra{0}e^{-i \tilde{K}t}\ket{0}=\sum_{m}\frac{\sqrt{2}}{2^{2 m}\sqrt{1+\sin \alpha}}\binom{2m}{m}(-1)^{m}\bigg(\frac{1-\sin \alpha}{1+\sin \alpha}\bigg)^{m}e^{-t/4}e^{- mt}\\=e^{-t/4}\sqrt{\frac{2}{1+\sin \alpha+e^{-t}(1-\sin \alpha)}}=\sqrt{\frac{1}{\cosh t/2+ \sin \alpha \sinh t/2}}
\end{gathered}
\end{equation}

\subsection{Matrix elements of the ordinary boost operator } 
\la{MatKApp}

Here we compute 
\begin{equation} \la{MatK}
    \bra{0}e^{-i K t}\ket{0}= \sum_\pm \int d\omega e^{-i\omega t}|\bra{\omega_\pm}\ket{0}|^{2}
\end{equation}
and we can evaluate the wavefunction overlaps in position space as
\begin{equation}
    \bra{0}\ket{\omega_\pm }=\sqrt{2}\int_{0}^{\infty} \frac{dx}{\sqrt{\pi}\pi^{1/4}} x^{-\frac{1}{2}+i 2\omega}e^{-x^{2}/2}=
    \frac{2^{i \omega}}{2^{1/4}\pi^{3/4}}\Gamma\bigg(i\omega+\frac{1}{4}\bigg)
\end{equation}
Inserting in \nref{MatK} leads to  \nref{OriKovrlp}. 
%
%
%
%

\subsection{Computations using coherent states} 
\la{CohApp}

The wavefunction of $\ket{\lambda}$ can be computed by writing $a^{\dagger}$ in terms of $p$ and $x$  
\begin{equation}
e^{\lambda a^{\dagger}}=e^{\lambda x/\sqrt{2}}e^{-i \lambda p/\sqrt{2}}e^{-\lambda^{2}/4}
\end{equation}
leading to 
\begin{equation}
\bra{x}\ket{\lambda}=\bra{x}e^{\lambda x/\sqrt{2}}e^{-i \lambda p/\sqrt{2}}\ket{0}e^{-\lambda^{2}/4}e^{-|\lambda|^{2}/2}=\frac{1}{\pi^{1/4}}e^{-|\lambda|^{2}/2}e^{-\lambda^{2}/2}e^{-\frac{1}{2}(x^{2}-2\sqrt{2}\lambda x)}
\end{equation}
Using that  
\begin{equation}
e^{-i K t}f(x)=e^{-t/4}f(x e^{-t/2})
\end{equation}
we get
\begin{equation}
\bra{x}e^{-i K t}\ket{\lambda}=\frac{e^{-t/4}}{\pi^{1/4}}e^{-|\lambda|^{2}/2}e^{-\lambda^{2}/2}e^{-\frac{1}{2}(x^{2}e^{-t}-2\sqrt{2}\lambda x e^{-t/2})}
\end{equation}
We now compute
\begin{equation}
\label{diaglamb}
\bra{\lambda}e^{-i K t}\ket{\lambda}=\int dx \bra{\lambda}\ket{x}\bra{x}e^{-i K t}\ket{\lambda}=\int dx \frac{e^{-t/4}}{\sqrt{\pi}}e^{-\frac{1}{2}(\lambda+\bar{\lambda})^{2}}e^{-\frac{1}{2}[x^{2}(1+e^{-t})-2\sqrt{2}(\lambda e^{-t/2}+\bar{\lambda})x]}
\end{equation}
one can then redefine $x=e^{t/4}y$ and integrate out $y$ to obtain 
\begin{equation} \la{MelK}
\bra{\lambda}e^{-i K t}\ket{\lambda}=\frac{1}{\sqrt{\cosh t}}e^{\frac{(\lambda e^{-t/4}+\bar{\lambda}e^{t/4})^{2}}{2\cosh t/2}-\frac{1}{2}(\lambda+\bar{\lambda})^{2}}
\end{equation}

\ 
We now turn to computations for the modified boost. We first note that 
\begin{equation} \la{creal}
a_{\alpha}=e^{\alpha P}a e^{-\alpha P}=\cos \frac{\alpha}{2}\, a+ \sin \frac{\alpha}{2}\, a^{\dagger} \text{, } e^{\alpha P}a^{\dagger} e^{-\alpha P}=\cos \frac{\alpha}{2}\, a^{\dagger}-\sin \frac{\alpha}{2}\, a
\end{equation}
We first compute many useful matrix elements. Consider the wavefunction 
$
\bra{x}e^{-\alpha P}\ket{\lambda}$. We observe that   $a_{-\alpha}e^{-\alpha P}\ket{\lambda}=\lambda \ket{\lambda}$. Using \nref{creal} this is a differential equation that implies 
%
\begin{equation}
\bra{x}e^{-\alpha P}\ket{\lambda}=\mathcal{N}_{\lambda,\alpha}e^{-\frac{1}{2}x \frac{[x(\cos \frac{\alpha}{2}-\sin \frac{\alpha}{2})-2 \sqrt{2}\lambda]}{(\cos \frac{\alpha}{2}+\sin \frac{\alpha}{2})}}
\end{equation}
We will also call  $\mathcal{N}_{0,\alpha}=\mathcal{N}_{\alpha}$.  We can solve for $\mathcal{N}_{\lambda,\alpha}$ in terms of ${\cal N}_\alpha$ by computing the following integral in two different ways
\begin{equation}
\bra{0}e^{-\alpha P}\ket{\lambda}=\int dx \bra{x}e^{-\alpha P}\ket{\lambda}\bra{0}\ket{x}=\int dx \bra{0}e^{-\alpha P}\ket{x} \bra{x}\ket{\lambda}
\end{equation} 
which can be seen to imply
\begin{equation}
\mathcal{N}_{\lambda,\alpha}=\mathcal{N}_{\alpha}e^{-|\lambda|^{2}/2}e^{-\frac{\lambda^{2}}{2}\big(\frac{1-\sin \alpha}{\cos \alpha}\big)}
\end{equation}
We can fix $\mathcal{N}_{a}$ by noting two things. First, we note that the quantity below is even in $\alpha$
\begin{equation} \la{Fsq}
\bra{0}e^{-\alpha P}\ket{0}=\pi^{-1/4}\mathcal{N}_{\alpha}\int dx e^{-\frac{\cos \frac{\alpha}{2} x^{2}}{\cos \frac{\alpha}{2}+\sin \frac{\alpha}{2}}}=\pi^{1/4} \sqrt{1+\tan \frac{\alpha}{2}}\mathcal{N}_{\alpha}=f(\alpha^{2})	
\end{equation}
It is even in $\alpha$ because the odd terms in $\alpha$ in the left hand side   either increase or decrease the oscillator level, having then zero overlap. The second thing to note is that
\begin{equation}
1=\bra{0}e^{-\alpha P}e^{\alpha P}\ket{0}=\mathcal{N}_{\alpha}\mathcal{N}_{-\alpha}\int dx e^{-\frac{x^{2}}{\cos \alpha }}=\sqrt{\pi \cos \alpha }\mathcal{N}_{\alpha}\mathcal{N}_{-\alpha}
\end{equation}
which, using \nref{Fsq} we get 
\begin{equation}
	\mathcal{N}_{\alpha}=\frac{\pi^{-1/4}}{\sqrt{\cos \frac{\alpha}{2}+\sin \frac{\alpha}{2}}}
\end{equation}	

This allows us to compute the matrix element that will be used in the norm calculation which is
\begin{equation}
\begin{gathered}
	\bra{0}e^{\alpha P}e^{-i K t}e^{-\alpha P}\ket{\lambda}=\\=\int \frac{dx}{\sqrt{\pi \cos \alpha}}e^{-\frac{\lambda^{2}}{2}\frac{1-\sin \alpha}{\cos \alpha}}e^{-\frac{x^{2}e^{t/2}}{2}\big(\frac{\cos \frac{\alpha}{2}+\sin \frac{\alpha}{2}}{\cos \frac{\alpha}{2}-\sin \frac{\alpha}{2}}\big)} e^{-\frac{x e^{-t/4}}{2}\frac{\big[x e^{-t/4} (\cos \frac{\alpha}{2}-\sin \frac{\alpha}{2})-2 \sqrt{2}\lambda \big]}{\cos \frac{\alpha}{2}+\sin \frac{\alpha}{2}}}
\end{gathered}	
\end{equation}
which one can simplify to 
\begin{equation}
	\bra{0}e^{\alpha P}e^{-i K t}e^{-\alpha P}\ket{\lambda}=\frac{1}{\sqrt{\cosh t/2+\sin \alpha \sinh t/2}}e^{-\frac{1}{2}|\lambda|^{2}}e^{-\frac{\sinh t/2 \cos \alpha}{\cosh t/2+\sin \alpha \sinh t/2}\frac{\lambda^{2}}{2}}
\end{equation}
so we can compute the norm by integrating the absolute value squared of the quantity above to find
\bea \la{NotKte}
 	|e^{-i \tilde{K}t}\ket{0}|^{2}&=&\frac{1}{\pi (\cosh t/2+\sin \alpha \sinh t/2)}\int d^{2}\lambda e^{-|\lambda^{2}|}e^{-\frac{\sinh t/2 \cos \alpha}{\cosh t/2+\sin \alpha \sinh t/2}\frac{(\lambda^{2}+\bar{\lambda}^{2})}{2}}\cr 
	&=&\frac{1}{\sqrt{1+\sinh t \sin \alpha+(\cosh t-1) \sin^{2}\alpha}}
\eea



\bibliographystyle{apsrev4-1long}
\bibliography{GeneralBibliography.bib}
\end{document}